\documentstyle[12pt]{article}
\advance\textheight by \topskip
\begin{document}
\begin{flushright}
NI94003\\
SU-ITP-94-9\\
hep-th/9407118
\end{flushright}
\vspace{-0.2cm}
\begin{center}
{\large\bf TOPOLOGY, ENTROPY AND  WITTEN INDEX \\
\vskip 0.5 cm
OF DILATON BLACK
HOLES}\\
\vskip 0.5 cm
  {\bf  G.W. Gibbons\footnote{Permanent address: D.A.M.T.P., Silver
Street,
University of Cambridge, England.\\
E-mail address: {\it G.W.Gibbons@amtp.cam.ac.uk} } and
R.E.  Kallosh\footnote{Permanent address:
Physics Department, Stanford University, Stanford   CA 94305, USA.\\
E-mail address: {\it kallosh@physics.stanford.edu}}}

Isaac Newton Institute for Mathematical Sciences,\\
University of Cambridge, 20 Clarkson Road, Cambridge, CB3 0EH, U.K.
\end{center}
\centerline{\bf ABSTRACT}
\begin{quotation}
 We have found that
 for extreme dilaton black
holes   an inner boundary
must be introduced in addition to the outer boundary to
 give an integer value to the Euler number.
The resulting manifolds have (if one identifies imaginary time)
topology
$S^1 \times R \times S^2 $ and Euler number
 $\chi = 0$ in contrast to the  non-extreme case with $\chi=2$.

The entropy of extreme $U(1)$ dilaton black holes is already known to
be
zero.
 We include a review of some recent ideas due to Hawking  on the
Reissner-Nordstr\"om
case. By regarding all extreme black holes as having an
inner boundary, we conclude that the entropy
of {\sl all} extreme black holes, including $[U(1)]^2$ black holes,
vanishes.

We  discuss the relevance of this to the vanishing of quantum
corrections
and
the idea that the functional integral for extreme holes gives
a Witten Index.  We have studied also the  topology of
``moduli space''  of multi black holes. The quantum mechanics on
black hole moduli spaces is expected to be supersymmetric despite the
fact
that
they are not
HyperK\"ahler since
the corresponding geometry
has   torsion unlike the  BPS monopole case.

Finally, we describe the possibility of extreme black hole fission
for states with an
energy gap. The energy released,  as a proportion of the initial rest mass,
during the decay of an
electro-magnetic
black  hole
 is 300 times greater than that released by the fission
of an
${}^{235} U$ nucleus.

\end{quotation}

\newpage
\section{Introduction}

There has been a great deal of interest recently in the classical
black hole solutions of theories with axion and dilaton fields
and the behaviour of such black holes in the quantum theory.
In particular there have been a number of new ideas about
pair creation, entropy and the evaporation of extreme black holes.
Our aim in this paper is to contribute to this discussion.
Because some of these new ideas may appear at first sight to be
puzzling,
if not paradoxical, we would like,
before passing on to the main business of the paper,
to make some general, and hopefully clarifying, remarks about the
significance
of classical solutions and what role one assigns them in the quantum
theory.

Traditionally one has thought of black holes, including primordial
black holes,
as having arisen from gravitational collapse. The classical solution,
and its attendant classical spacetime
is expected to provide a good description for holes rather larger
than the
Planck mass, but as the black hole losses mass by Hawking evaporation
one
expects that the classical picture should become  less and less
accurate.
At this stage one might anticipate that processes that
are classically forbidden,  such as fission, might become important.

If
the initial black hole has a charge which because either, (i) no
ordinary particles carry it, or (ii) if they do they are so heavy as
to
suppress
its creation by the field of the hole, then the Hawking evaporation
is expected to lead to a limiting, near extreme state. Actually in
case (i)
it is difficult to see how the hole can have acquired the charge in
the first
place
unless it was created in a pair creation process.
This latter process is inherently quantum mechanical. One might be
tempted
to describe it using instanton methods and again a classical solution
is
involved. However it is not immediately clear that the appropriate
boundary
conditions
at the horizons of {\sl these}  classical solutions are the same as
those that apply to classical macroscopic black holes.
This is particularly true in the extreme case. In the classical
theory one never encounters an extreme black hole as such -- one
merely finds that under suitable circumstances that a sub-extreme
hole may become more and more extreme, rather like a (massive)
particle which may become more and more relativistic but which never
actually
attains the velocity of light. However, just as with massless
particles, it is
possible to contemplate a class of extreme black holes in their own
right.
Presumably such extreme black holes may {\sl only} be
produced by pair creation. If this is true then some properties of
extreme
holes might indeed be different from near extreme holes.
Superficially this
might give rise to the impression that some physical properties
may change discontinuously as the extremal limit is approached.
However
such an impression would be illusory if the limit, as is suggested
here,
is never attained.

A second role for classical solutions in a field theory is as
``solitons''. Solitons are
 typically topological non-trivial
classical field configurations
carrying a conserved charge -- often a ``central'' charge  in a
supersymmetric
theory -- which is not carried by the  ``elementary''  fields. There
are many reasons for regarding extreme black holes as the
solitons
of  supergravity and superstring theories. One often thinks of
soliton
as a weak coupling approximation to a single quantum state in the
underlying
Hilbert space of the theory. If extreme holes are solitons it seems
reasonable
to expect that:
\begin{enumerate}
\item Classical extreme hole solutions should have a different
topology
from non-extreme holes.
\item  The classical entropy of the soliton solutions should vanish.
\end{enumerate}
It is also not unreasonable to expect that soliton solutions might
satisfy
different boundary conditions from solutions representing non-extreme
black holes formed during gravitational collapse, since solitons are
usually
only formed by pair creation.

A third role for classical solutions is as saddle points in the
approximate evaluation of the functional integral  giving some
matrix element
or giving
a partition function. The classical solutions are typically
``euclidean'',
i.e. have signature ++++ and the appropriate boundary conditions then
depend upon what matrix element or what partition function one is
trying to
evaluate. Thus if one wants to evaluate the thermodynamic properties
of black holes one sums over bosonic fields which are periodic in
imaginary
time.
Naturally the classical bosonic saddle point is periodic in imaginary
time. This technique may be used to calculate the entropy of a system
(containing black holes) at fixed temperature $T= \beta^{-1}$. In our
opinion a
great deal of wasted effort has gone into
to trying to interpret this equilibrium too literally in a purely
classical way.
Classically such an equilibrium is impossible. The classical
solutions are
saddle points used in the evaluation of the functional integral. They
do not
represent familiar astrophysical black holes. The classical saddle
points
are, for example, time symmetric and have both past and future
horizons.
Put more picturesquely they contain both black holes and white holes.
One does not expect to encounter white holes in the classical theory.

In any event one finds in this way that non-extreme classical black
holes
contribute an amount S of entropy given by
\begin{equation}
S= { 1 \over 4} A\ ,
\label{entr}\end{equation}
where $A$ is the area of the event horizon. Quantum fluctuations
will,
in general, modify this formula. The classical saddle point giving
this
result has its mass and charge  determined by the temperature and
potential of the system.

To what extent one can trust eq. (\ref{entr})
for
charged black holes? For the  Schwarzschild black hole  the
thermodynamical
approach obviously breaks down when the mass of a black hole becomes
smaller
than  the Planck mass, and its temperature becomes greater than the
Planck
mass. For the
charged black holes the temperature reaches some maximum value and
then
approaches zero in the
extreme limit. Thus one might expect that the semiclassical
description of the
charged black holes to be quite reliable. However, a more detailed
investigation
of
this question performed in \cite{PW,US} casts doubt on this.
The standard description breaks down if the  emission of a
single particle with the energy of the order of Hawking temperature
would make
the mass of the black hole smaller than its extreme value.
Starting from this
moment the
thermodynamical approach becomes ambiguous; it does not allow one to
calculate the entropy of the near extreme black holes in a reliable
way, since
one
can no longer neglect the backreaction of created particles.

The situation with respect to extreme static holes is even
more complicated.
Firstly if the coupling $a$ to the dilaton is non-zero then the
metric of $U(1)$
black holes
is singular on the horizon (at least in Einstein conformal gauge).
Secondly the area of the event horizon of these black holes tends to
zero as the extreme
limit is approached. This presumably means that their contribution
to the entropy vanishes.  One of the purposes of the present paper is
to check
this point in detail. If the coupling to the dilaton vanishes, i.e.
in the
Einstein-Maxwell case, the situation is slightly different but the
end result is
similar. The evaluation of the action requires additional inner
boundary terms \cite {HHR}. These are chosen to make the extreme
holes a {\it bona fide}
solution of the classical variational problem. These boundary terms
contribute to the classical  action and hence to the classical
entropy
of the
Reissner-Nordstr\"om black holes
with the result that it vanishes \cite {HHR}, see also \cite {T}:
\begin{equation}
S=0\ .
\end{equation}
Of course, in general, thermal fluctuations and closed loops might be
expected
 to give a non-vanishing contribution. The latter will in general
diverge
except in a well defined quantum theory such as superstring theory
might be.
One would not, however, expect the entire contribution to vanish
because
thermal fluctuations should contribute differently and moreover
positively
to the entropy, be they bosons or fermions.

In addition to the thermal or Gibbs partition function $Z(\beta) =Tr
e^{ -
\beta H}$ in a supersymmetric theory, one  may be interested in the
Witten index $Y(\beta)= Tr (-)^F e^{- \beta H}$. This is given
by a functional integral over boson and fermion fields {\sl both}
 of which are periodic in imaginary time. Solitons and  extreme
holes holes should contribute to $Y(\beta)$. Non-extreme holes
cannot.
This is because fermion fields must,
for purely topological reasons, be {\sl anti-periodic} on these
backgrounds \cite{GibbonsPerry}. The point here is that
for topology $S^2 \times R^2$ the spin structure is unique
because the manifold is simply connected.
Moreover this unique spin structure
corresponds to spinors which are antiperiodic at infinity.
For extreme holes we  will argue that the topology is $S^1 \times R
\times
S^2$ which
is not simply connected. Two spin structures are allowed,
one of which is periodic at infinity.

 Being a topological quantity, quantum and thermal
fluctuations of fermions and bosons should
cancel in the expression for $Y(\beta)$. For this reason it may make
more
sense to evaluate $Y(\beta)$ in the field theory limit of a
putatively finite
superstring theory. To do so one would need to consider boundary
conditions on the extreme classical background in imaginary time
which
are quite different from what one uses for non-extreme holes.
This is one of the principal points we wish to make in this paper.

We have given three different ways in which classical black hole
solutions may be used in the quantum theory. No doubt the reader can
think of
others. Each forms the basis of a perturbative calculation of some
process or
physical property of a quantum black hole.
In effect however nobody  knows what a quantum black hole is really
like,
or indeed whether such a concept is well defined in the full quantum
theory --
not  least, because we don't yet have the full
quantum theory,
let alone the tools needed to analyse it.
It may be that some special theories, heterotic string theory for
example,
exhibit an Olive-Montonen type  duality  which makes statements about
solitons in the weak coupling limit exact statements about string
states
in the strong coupling limit. In that case the classical concept
of an extreme black hole may carry over completely into the quantum
regime.
Even then it is not obvious that the detailed properties
 will have much
similarity
with what we expect of macroscopic astrophysical black holes.
In theories which do not admit such a duality one may have to
 recognize
that the idea of a black hole is essentially a classical concept and
that one
has no right to be dogmatic ahead of time about what modification may
be
necessary in the quantum theory. Thus for example,
 the quasi-teleological properties ascribed in \cite {HHR} to
extreme black
holes, while unfamiliar, may indeed be appropriate for particular
calculations,
for example  of the Witten index. It is neither obvious nor
excluded that they are appropriate for some sorts of scattering
calculations.
For that reason we have tried to explore some of the geometrical
properties of
extreme black holes and their possible physical consequences but any
such
consequences do not follow directly from the solutions themselves but
rather,
as we have tried to indicate above, from a whole set of physical
assumptions
about the physical role of the classical solutions in a particular
calculation,
subject to appropriate boundary conditions. Strictly speaking,
statements
such as ``black holes have such and such properties''
are meaningless without these implicit assumptions being
carefully spelt out. In the classical theory considerable agreement
exists
about what the appropriate  physical assumptions  are. In the quantum
theory no
such agreement is  even possible in principle at present because we
lack such a
theory.
For that reason, and others, we have declined in this paper to
speculate
extensively about the possible consequences of our results for black
hole
evaporation and the so-called information paradox.
However this process operates it is certainly a {\sl dynamical}
 process and we feel that only a limited insight can be gained by
examining the
properties of purely {\sl static} classical solutions.
This is not the least so because the classical solutions  may
have past and
future event horizons (or naked singularities). Classically these
render the
classical evolution ambiguous. Presumably this ambiguity would be
resolved in a
proper, dynamical quantum mechanical treatment. This we reserve for
future
study.

To summarize: axion-dilaton black holes are
different from those of Schwarzschild, Reissner-Nordstr\"om and
Kerr-Newman black holes.  The presence of dilaton has a great affect
on
the geometry as well as the thermodynamical properties of these
objects.
Some properties of dilaton black holes are exactly the same as
those of
classical ones and some are not. For example,   the entropy $S$
calculated
by the standard Euclidean methods is always
equal to
one quarter of the area of the horizon. However, Schwarzschild,
 and Kerr black holes never have vanishing surface area
and hence entropy.
By contrast some extreme dilaton black holes  do have vanishing area
of the
horizon and
 vanishing entropy. The study of the topology and the entropy of the
charged
 black holes and specifically of the charged
extreme black holes  is the purpose of this paper. In fact we shall
only
treat in detail static black holes and we have at present nothing
new to say about the extreme Kerr solution.

The  Euler number of the Schwarzschild and Kerr
black holes equals 2  since the corresponding manifold is identified
to be  $R^2 \times S^2$,  see Fig. 1. The Gauss-Bonnet (G-B) theorem
confirms this by
representing this number as the volume G-B
integral supplemented by the outer boundary term \cite{GH,EGH}.
One motivation for the present work was whether this going to be the
same for
all charged dilaton black holes and what happens in the extreme
limit.

Because in the Einstein conformal gauge extreme holes
may be singular at the horizon and so standard definitions may break
down,
in what follows we are going to assume
that by definition the extreme solution is the one which allows one
to identify the imaginary time coordinate $\tau = it$ with {\sl any}
period
$\beta$. One is led to the following picture. For extreme black holes
the
topology of
Riemannian section is $S^1 \times R \times  S^2$ and hence the Euler
number should vanish, $\chi= 0$, see Fig. 2.

Thus our first aim is to understand better
the topology and in addition to calculate
the  Euler number of all charged axion-dilaton black holes using
the explicit form of the metric in the Euclidean space and the G-B
theorem.
The second reason why we are interested in Gauss-Bonnet
Lagrangian is
related to the  trace anomalies in gravity.
The 4-dimensional trace anomaly contains (in addition to other
terms) the
following expression \cite{CD}.
\begin{equation}
T(x) =  g_{\mu\nu} < T^{\mu\nu}> = {A\over 32 \pi^2} {}^*
R_{\mu\nu\lambda\delta} {}^*  R^{\mu\nu\lambda\delta} \ .
\end{equation}
The coefficient $A$ is known for all fields interacting with gravity.
The integrated form of the anomaly in the Euclidean space expresses
the trace
of the energy-momentum tensor  through the Euler
 number of a
manifold
\begin{equation}
\int d^4x \sqrt{-g} \;T(x)  = A \; \chi \ .
\end{equation}
The knowledge of the local expression for the G-B term may be
useful for the analysis of the way in which the quantum anomaly
affects the
classical
solutions.
The third reason for our interest to G-B term is the following. The
above mentioned anomaly is actually related to the fact that in
topologically
non-trivial backgrounds there exists a counterterm, proportional to
the G-B
action. Therefore in general one may want to consider the classical
gravitational action supplemented by a G-B term with some arbitrary
coefficient
in front of it.\footnote{This is in an analogy with the well known
fact that
the  pure Yukawa
coupling $g\bar \psi \psi \phi$ in quantum field theory without
$\lambda \phi^4$ coupling is inconsistent.
Only in
presence of
non-vanishing $\lambda$ the theory is consistent as a renormalizable
quantum
field theory. The actual value of both couplings is determined by the
flow of
renormalization group but it is inconsistent to take $\lambda=0$ from
the beginning.
Of course,  gravity is not renormalizable but it still seems logical
to proceed
as in field theory when dealing with one-loop divergences. }
This action will be able to
absorb a corresponding one-loop divergence. This in turn may affect
any
 calculation of the action for some configurations. Thus we would
like
to
understand the value of the G-B action for the new family of dilaton
black
holes, including extreme ones.  The presence of
such term may  affect different configurations in a different way.

Our interest to the problem was greatly enhanced by the recent
calculations, following an earlier suggestion \cite{GIBB},
 of the rate of the pair production of extreme black holes
\cite{Fay,Ross}.
This
process  involves change in topology, and therefore the one-loop
Gauss-Bonnet
correction to the classical Einstein-Maxwell-dilaton-axion action
may
become relevant.
In addition, it was observed in \cite{Fay}, that   the calculation of
the
instanton action related to the pair production of
Reissner-Nordstr\"om black
holes has a strange feature: discontinuity at the limit to extreme.
This leads
to a prediction that
the rate of production of extreme black holes is suppressed
comparatively to the
non-extreme ones by the amount equal to the entropy of the extreme
Reissner-Nordstr\"om black holes.  The physical meaning of this
discontinuity
is the subject of a recent paper by Hawking, Horowitz and Ross \cite
{HHR}.

 In addition the mere existence of instantons,
describing pair
production of non-extreme $U(1)$ black holes depends on  the
dilaton coupling; in fact they exist only for  $0 \leq a<1$ although
the
instantons
for pair
production of
extreme black holes do exist for stringy case $a=1$.
The instantons describing pair production of  $[U(1)]^2$
non-extreme as
well as extreme black holes have been constructed by Ross
\cite{Ross}. The area of
the horizon of these extreme black holes is not vanishing.
We will show that the inner boundary at the horizon, which will be
introduced for all
extreme black holes, will change the value of their entropy as
dramatic as for
Reissner-Nordstr\"om black holes. Instead of being
\begin{equation}
S_{[U(1)]^2}={A\over 4} = \pi (M^2- \Sigma^2)\ ,
\label{entr2}\end{equation}
 where $\Sigma$ is the dilaton charge,
the entropy will vanish. In the Reissner-Nordstr\"om case $\Sigma=0$.

Another closely related issue is the possibility that at the last
stages of the
black hole evaporation, when black hole approaches its extreme limit,
 a
black hole may continue evaporating by
splitting into smaller extreme black holes carrying elementary
electric and
magnetic charges \cite{US}.  As  was argued in \cite{US}, the
probability of
such process should be suppressed by the factor $\exp (\Delta S)$,
where
$\Delta S$ is the change of entropy.  For large black holes the
change of
entropy $\Delta S$ is large and negative, which prevents their
splitting, in
accordance with the  second law  of black hole physics. Meanwhile,
some extreme
black holes studied in \cite{US} have vanishing entropy, which allows
them to
split without violation of the  second law  of black hole physics. In
this
paper we will argue that in fact {\it all} extreme dilaton black
holes have
vanishing entropy. This makes the possibility of splitting of extreme
black
holes  more plausible.

Finally the vanishing Euler
 number and the vanishing entropy
seem  an essential component of a new interpretation of the
functional
integral for extreme black holes in supersymmetric theories
as being purely topological in nature and giving a Witten Index
for supersymmetric soliton states. In fact
independently of that interpretation one might argue  that a zero is
the only
consistent value for the entropy of  solitons.

Thus we have presented various reasons why the investigation of the
topology of
all known axion-dilaton black holes and of  the behavior of the
Gauss-Bonnet
action is worthy of study.

The plan of the paper is the following.  In Sec. 2 we
will present a simple proof that the 4-dimensional G-B volume
integral
over the
compact manifold without a boundary is a topological invariant.
The next step is the calculation of the G-B Lagrangian for the static
spherically symmetric non-extreme
Euclidean axion-dilaton black holes. First we calculate the volume
integral in
Sec. 3.
In Sec. 4 the general discussion of the boundary terms for the G-B
action is
presented,
which is based on the work of Chern \cite{C}. We discuss a special
situation for
manifolds with boundary, when the Euler number vanishes.
In Sec. 5
we calculate the volume
G-B integral over a
compact manifold by putting
the outer boundary at some radius $r_0$.  The boundary terms at this
 boundary are calculated, which finally allows to verify the G-B
theorem
for all non-extreme black holes. The fact that the Hawking
temperature is
finite and non-vanishing plays an important role in providing a
compactness
of our manifold in the Euclidean time direction.
Under specific assumptions about the area of integration in the
Euclidean
time
we proceed with the corresponding calculations for the extreme black
holes in Sec. 6. We will come to the conclusion that without placing
an
additional
inner boundary we cannot have a consistent picture for extreme black
holes.
The existence of this inner boundary affects the calculation of the
entropy of extreme black holes, which is presented in Sec. 7. In Sec.
8 we
define
 the Witten Index of the fluctuations existing in the extreme black
hole
background. The supersymmetric non-renormalization theorem which was
known before for the entropy of such black holes becomes relevant for
the
topological Witten Index of the solitons. In Sec. 9 we prove that
the  Euler number of Euclidean multi black hole solutions
 vanishes. The topology of the ``moduli space''  of
multi black holes is studied both by standard methods for the compact
spaces as
well as by investigation of the corresponding de Rahm complex.
In Sec. 10 an example of  $(4,1)$ supersymmetric sigma model is given
by the
target space of an uplifted dilatonic non-Abelian magnetic black
hole.
The non-HyperK\"ahlerian properties of the  manifold are due to the
existence of
the torsion in the uplifted geometry of the magnetic black hole with
the unbroken
 space-time supersymmetry.  In Sec. 11 we have suggested to consider an
one-dimensional supersymmetric action in the black hole background to
study  the supersymmetric quantum mechanics of the extreme black holes.
Sec. 12 describes various examples of
splitting and fission of extreme black holes.

Some details of Gauss-Bonnet
calculations can be found in Appendix.  The figures illustrate the
topology of
non-extreme and extreme black holes and their entropy and the
topology of the
\lq \lq
relative moduli space\rq \rq of two extreme black holes.

\section{Topological invariance}
We will start with a short reminder of the
topological character of the G-B action \cite{C,EGH}.  The Euler
 number
of a
4-manifold is the alternating sum of the Betti numbers,
\begin{equation}
\chi [{\bf {\cal M}}] = \sum_{p=0}^{p=4} (-1)^p B_p\ .
\end{equation}
Gauss-Bonnet theorem relates the Euler
 number of the closed Riemannian
manifold without a boundary to the
volume integral of the curvature of the four-dimensional metric,
\begin{equation}
S_{GB}= {1\over 32 \pi^2}\int_M
\epsilon_{abcd}\; R^{ab} \wedge R^{cd} \ .
\label{1}\end{equation}
In equation (\ref{1}) the curvature 2-form is defined as
\begin{equation}
R^a{}_b = d \omega^a{}_b + \omega^a{}_c \wedge \omega^c{}_b
\ ,
\end{equation}
where $\omega^a{}_b$ is a spin connection one-form.
Consider some arbitrary variation of the metric, which will result in
some
variation of the spin connection $\delta \omega^a{}_b$. Under such
variation
the volume integral in eq. (\ref{1}) changes as follows:
\begin{equation}
\delta S_{GB}=  {1\over 32 \pi^2}\int_M  2 \; \epsilon_{abcd}\;
\delta R^{ab} \wedge R^{cd} \ ,
\end{equation}
where
\begin{equation}
\delta R^a{}_b = d\;  \delta \omega^a{}_b + \delta \omega^a{}_e
\wedge
\omega^e{}_b + \omega^a{}_e \wedge  \delta \omega^e{}_b = D \;
\delta \omega^e{}_b \ .
\end{equation}
We get
\begin{equation}
\delta S_{GB}=  {1\over 32 \pi^2}\int_M  2 \epsilon_{abcd}\;
 (D\;  \delta \omega^{ab})  \wedge
 R^{cd} \ .
\label{var}\end{equation}
Using the Bianchi identity for the Riemannian manifold
\begin{equation}
D R^a{}_b = 0
\end{equation}
and the covariant constancy of the tensor $\epsilon_{abcd}$ in the
form
\begin{equation}
D \epsilon_{abcd}= 0
\end{equation}
we can bring equation (\ref{var}) to the form
\begin{equation}
\delta S_{GB} =  {1\over 16 \pi^2}\int_M   \;D\;
(\epsilon_{abcd}\;
  \delta \omega^{ab}  \wedge
 R^{cd}) \ .
\end{equation}
Finally, we may notice that our covariant derivative  $D$ acts on an
$SO (4)$
invariant differential form  and therefore we may replace it by the
ordinary derivative $d$, acting on this 3-form. Thus the volume part
of the
G-B action under the arbitrary change of the metric undergoes the
following
change:
\begin{equation}
\delta S_{GB} =  {1\over 16 \pi^2}\int_M   \;d\;
(\epsilon_{abcd}\;
  \delta \omega^{ab}  \wedge
 R^{cd})  \ .
\end{equation}
We have shown that the variation of the Gauss-Bonnet form is an exact
form. If
the manifold has no boundary, the integral from the exact form
vanishes
according to the Stoke's theorem.  This proves the topological
character of
this expression, specifically its independence of the metric. If
however the
manifold does have a boundary, the volume integral (\ref{1})  does
depend on
the metric since the variation equals to
\begin{equation}
\delta S_{GB}^{volume} =  {1\over 16 \pi^2}\int_{\partial M }
(\epsilon_{abcd}\;
  \delta \omega^{ab}  \wedge
 R^{cd})  \ .
\end{equation}
The explicit expressions for the boundary terms for the Euler
 number
have been
presented  in \cite{GH} and \cite{EGH} in a different form.

\section{G-B volume integral}

 Now we can start calculating
the G-B Lagrangian for axion-dilaton dilaton black holes
\cite{GM,US,axion}.
The metric of spherically
symmetric  $U(1)$ and
$U(1)\times U(1)$ black holes is given by
\begin{equation}
ds^2 = - e^{2U}(r) dt^2 + e^{-2U}(r) dr^2 + R^2 (r) d^2 \Omega \ .
\label{met}\end{equation}
The G-B volume integral
\begin{equation}
 S_{GB}^{volume} = {-1\over 32 \pi^2}\left[\int_M
\epsilon_{abcd}\; R^{ab} \wedge R^{cd}\right] =  {1\over 32
\pi^2}\left[\int
d^4 x \sqrt{-g} \;(R_{\mu\nu\lambda \delta} R^{\mu\nu\lambda \delta}
-4 R_{\mu\nu} R^{\mu\nu} + R^2)
\right]
\label{GB1}\end{equation}
can be calculated using the values of the Riemann
tensor $R_{\mu\nu\lambda \delta}$,
Ricci tensor $R_{\mu\nu}$ and Ricci scalar $R$ for the metric of the
form
(\ref{met}). The result of such calculation is the following,  see
Appendix:
\begin{equation}
\sqrt{-g} \;(R_{\mu\nu\lambda \delta} R^{\mu\nu\lambda \delta}
-4 R_{\mu\nu} R^{\mu\nu} + R^2) = {\partial \over \partial r}\left[4
(e^{2U})'
\left(1 - (e^{U} R')^2\right) \right]  \ .
\label{append}\end{equation}
The volume integral takes the form
\begin{equation}
S_{GB}^{volume} = {1\over 32 \pi^2}\left[\int
d^2 \Omega \;dt\; dr {\partial \over \partial r}\left[4 (e^{2U})'
\left(1 - (e^{U} R')^2\right) \right]\right]\ .
\end{equation}
It shows indeed that the G-B integrand is a total derivative. For
this class of
metrics we may perform the angular integration and we are left with
\begin{equation}
S_{GB}^{volume} = {-1\over 2 \pi}\left[\int
dt\; dr {\partial \over \partial r}\left[  (e^{2U})'
\left(1 - (e^{U} R')^2\right) \right] \right] \ .
\end{equation}
If we work in a Lorentzian spacetime with  Minkowski signature there
is nothing
interesting to say about
the
integral. The integrand is time independent and the range of
integration in
$t$ is infinite. One could identify in real time but this would
introduce closed timelike curves and would also lead to
singularities on the horizon.  However if  we use the Riemannian
version of the
metric,
\begin{equation}
ds^2 =  e^{2U}(r) dt^2 + e^{-2U}(r) dr^2 + R^2 (r) d^2 \Omega  \ ,
\label{euclmet}\end{equation}
where $\tau$ is a periodic coordinate, the range of
$\tau$-integration is
constrained from $0$ to $ \beta$ by the standard requirement
\cite{H}
\begin{equation}
\beta  = {2\pi \over \kappa}  \ .
\end{equation}
This range of integration is finite and not vanishing under the
condition that
the surface gravity, which is given by
\begin{equation}\label{surgrav}
\kappa = \frac{1}{2} { \frac{\partial_r g_{tt}}{\sqrt{-g_{rr}
g_{tt}}} }
\biggm|_{r=r_h} \  ,
\end{equation}
is finite.  For such class of metrics one can find a change of
coordinates
where the apparent singularity at the horizon at $e^{2U}(r= r_{h})
=0$ is
removed.

 When the surface gravity does vanish, as for some extreme black
holes with zero temperature, the range of integration in $\tau$ is
infinite.
When the surface gravity is infinite, as for some  extreme black
holes with
$a>1$, the range of integration in
$\tau$ has to shrink to zero to give a  finite result. We do not at
present
see how to give a straightforward meaning to the G-B
integral for such solutions and will limit ourselves at present   to
the $a\leq
1$ solutions
only.

Thus we consider a  class of metric (\ref{euclmet}) and the Euclidean
time range is  $0 \leq \tau \leq \beta$, where
\begin{equation}
\beta = 4\pi \left((e^{2U})'|_{r=r_h}\right)^{-1}  \ .
\label{beta}\end{equation}
 We will consider first the non-extreme black holes, when
this range is finite, i.e. the temperature of the black hole
$T= 2\pi \kappa$ is finite,
and non-vanishing. There is a regular horizon, and we may
integrate over time and get
\begin{equation}
S_{GB}^{volume} = - 2  \left((e^{2U})'|_{r=r_h}\right)^{-1}
\left[\int_{r_h}^{r_0}
 dr {\partial \over \partial r}(e^{2U})'
\left(1 - (e^{U} R')^2\right)
\right]\ .
\end{equation}
We consider a compact manifold with $r_h \leq r \leq r_0$.
The $r$-integration may also be performed and the result is
\begin{equation}
S_{GB}^{volume} =  2  (e^{2U})'|_{r=r_h})^{-1} \left\{ \left[
(e^{2U})'
\left(1 - (e^{U} R')^2\right)
\right]_{r=r_h} - \left[
(e^{2U})'
\left(1 - (e^{U} R')^2\right)
\right]_{r = r_0}   \right\} \ .
 \end{equation}

Thus the volume G-B integral for the black holes with regular event
horizon is
\begin{equation}
S_{GB}^{volume} =  2
\left(1 - (e^{U} R')^2\right)
|_{r=r_h} -  2  (e^{2U})'|_{r=r_h})^{-1} \left[
(e^{2U})'
\left(1 - (e^{U} R')^2\right)
\right]_{r = r_0}  \ .
\label{vol}\end{equation}

The second term in this equation  is vanishing in the limit when $r_0
\rightarrow
\infty$ for black holes under consideration, since they are
asymptotically flat
spaces.

However, the G-B theorem applies only  to compact manifolds perhaps
with
boundaries, which
means that we cannot consider such limit. We have
to comply with the conditions of the theorem and consider a manifold
with a
boundary at $r=r_0$.

\section {Boundary corrections}
To treat the G-B  action with boundary terms it is convenient to use
differential forms.  Chern \cite{C} has considered  the Gauss-Bonnet
form
$\Omega $ of degree  $n=2p$,
\begin{equation}
\Omega = {(-1)^p\over 2^{2p}  \pi^p p!} \epsilon_{a_1 \dots a_{2p} }
R^{a_1a_2}
\wedge \dots  R^{a_{2p-1} a_{2p}}  \ .
\label{CH}\end{equation}
He has shown that this form, which is originally defined in a closed
manifold
$M^n$ of $n$ dimensions, can be defined in a
manifold $M^{2n-1}$ of dimension $2n-1$. This larger manifold is
formed by the
unit vectors of the original manifold.  Chern has shown that $\Omega$
is equal
to the exterior derivative of a differential form $\Pi$ of degree
$n-1$ in
$M^{2n-1}$,
\begin{equation}
\Omega = - d \Pi  \ .
\label{CHERN}\end{equation}
The original integral over $M^n$ of the form $\Omega$ is equal to the
same
integral over a submanifold $V^n$ of $M^{2n-1}$, and, according to
the Stoke's
theorem, to the integral over the boundary of $V^n$ of $\Pi$. Since
the unit
vectors have some isolated singular points, the boundary of $V^n$
corresponds
to the singular points of the vector field defined in $M^n$. The
integral of
$\Pi$ over the boundary of $V^n$
is evaluated and proved to be equal to $\chi$. Thus according to
Chern we get
\begin{equation}
S_{GB}^{volume} =\int_{M^{n}}\Omega = \int_{V^{n}} \Omega =
   \int_{\partial V}  \Pi \ .
\label{chern}\end{equation}
For four-dimensional case the boundary terms have been given in
explicit form
using differential forms in \cite{EGH}. Actually by comparing their
boundary
terms with the  expression given by Chern \cite{C} for $\Pi$ one may
notice
that
EGH expression for the Euler
number presented in eq. (\ref{GB}) can be
rewritten as
\begin{equation}\label{GB}
S_{GB}= S_{GB}^{volume} + S_{GB}^{boundary}=\int_M  \Omega
  - \int_{\partial M}  \Pi  = \int_{\partial V}  \Pi  -
\int_{\partial M}  \Pi
\ .
\end{equation}
This expression
provides an exact cancellation of the volume term by the boundary
term
under condition that the boundaries of the manifold $M^n$ and of a
submanifold $V^n$ of the manifold $M^{2n-1}$ are the same.

For non-extreme black holes, which have regular horizon  in Minkowski
space and
finite temperature, the Euler
number equals 2, as explained in Sec. 2. Let us show how the G-B
theorem
reproduces this result. Equation (\ref{GB})  tells us that if the
boundary of a
submanifold $V^4$  is at  $r=r_0$ as well as at $r= r_h$ we should
get
\begin{equation}
S_{GB}= S_{GB}^{volume} + S_{GB}^{boundary}= \int_{r_0}  \Pi  -
\int_{r_h}  \Pi
 - \int_{r_0}  \Pi =  -\int_{r_h}  \Pi
 \ .
\label{GBnonextreme}\end{equation}

This is equivalent to the statement that the contribution from the
volume
integral at the outer boundary is exactly compensated by the boundary
term, if
the boundary term
is correct. Basically it is a statement that  there is  no need to
calculate
it, one has to
ignore the contribution from $r_0$ when calculating the volume
integral, as if
it would be legal to take the limit $r_0 \rightarrow  \infty$.

Alternatively rather than using the papers by  Chern \cite{C}  we
could use
the G-B theorem as given in \cite{EGH}, where the boundary terms are
written
explicitly.
\begin{equation}
S_{EGH}^{bound}= -{1\over 32 \pi^2}  \int_{\partial M}
\epsilon_{abcd}
\; (2 \theta ^{ab} \wedge R^{cd}- {4\over 3} \; \theta ^{ab}\wedge
\theta
^{a}{}_e\wedge \theta ^{eb}) \ .
\label{b}\end{equation}
In equation (\ref{b}) $\theta^{ab}$ is the second fundamental form of
the
boundary.
Let us  specify all the above-mentioned forms for the axion-dilaton
black holes
under consideration.
Our metric (\ref{euclmet}) corresponds to the following
vierbein forms
\begin{equation}
e^0 = e^{U}(r) dt,\quad e^1 = e^{-U}(r) dr, \quad e^2 = R(r) d\theta,
\quad e^3
= R(r)
\sin\theta d\phi  \ .
\end{equation}
The spin-connections are
\begin{equation}
\omega^{01} =1/2 (e^{2U})' dt, \quad \omega^{21} = e^U R' d\theta,
\quad
\omega^{31} = e^U R'\sin\theta d\phi, \quad \omega^{32} = \cos\theta
d\phi\ .
\end{equation}
The curvature  2-forms is given by $R= d\omega +
\omega\wedge \omega$, see Appendix.

The volume G-B integral for this curvature becomes
\begin{equation}
S_{GB}^{volume} = {1\over 32 \pi^2}\int_M  \epsilon_{abcd} R^{ab}
\wedge
R^{cd}= {1\over 4 \pi^2}\int_V  d \left(\omega^{01} \wedge
R^{23}\right) \ ,
\end{equation}
where \begin{equation}
R^{23} = d\omega^{23} + \omega^{21}\wedge \omega^{13}= \sin \theta
\left(1 - (e^U R')^2\right)  d\theta  d\phi  \ .
\end{equation}
Using Stoke's theorem we get
\begin{equation}
S_{GB}^{volume} =  {1\over 4 \pi^2}\int_{\partial V }\omega^{01}
\wedge R^{23}  \ .
\end{equation}
Our assumption about the range of integration in Euclidean time,
given in eq.
(\ref{beta}) in terms of the  integral of the one-form is reduced to
\begin{equation}
{1\over 2 \pi} \int \omega^{01}|_h ={1\over 2 \pi}  \int _0^{\beta}
d\tau \;
\omega_{\tau} ^{01}|_h = \int  _0^{\beta}  d\tau  \; T = 1\ ,
\label{cond}\end{equation}
where $T$ is the black hole temperature.
Integrating out the angular variables and considering the boundary
$\partial V$
to be at $r_0$ and at $r_h$ we have as before
\begin{equation}
S_{GB}^{volume} =  2   \left(1 - (e^{U} R')^2 \right)|_{r_h}    -  2
(e^{2U})'|_{r=r_h})^{-1} \left[
(e^{2U})'
\left(1 - (e^{U} R')^2\right)
\right]_{r = r_0}  \ .
\end{equation}
To specify the boundary corrections we may define the metric at the
boundary
$r=r_0$ as
\begin{equation}
ds^2_0 =  e^{2U}(r_0) dt^2 + e^{-2U}(r_0) dr^2 + R^2 (r_0) d^2 \Omega
 \ .
\label{bounmet}\end{equation}
The non-vanishing spin connections of this metric are
$(\omega^{32})_0 =
\cos\theta\, d\phi$. Hence the second fundamental form
$\theta^{ab} = \omega^{ab} - (\omega^{ab})_0$ at the boundary
$r=r_0$ is
\begin{equation}
\theta^{01} =1/2 (e^{U})' dt, \quad \theta^{21} = e^U R' d\theta,
\quad
\theta^{31} = e^U R'\sin\theta d\phi, \quad \theta^{32} = 0\ .
\label{second}\end{equation}
The calculation of the EGH boundary term
\begin{equation}
S_{EGH}^{bound}= -{1\over 32 \pi^2}  \int_{\partial M}
\epsilon_{abcd}
\; (2 \theta ^{ab} \wedge R^{cd}- {4\over 3} \; \theta ^{ab}\wedge
\theta
^{a}{}_e\wedge \theta ^{eb})
\label{bound}\end{equation}
is straighforward.
We find that the part of the first term in eq.
(\ref{bound}) which does not contain $\theta^{01}$ is completely
cancelled
by the second term. The part which does contain $\theta^{01}$ is
\begin{equation}
S_{EGH}^{bound}= -{1\over 4 \pi^2}  \int_{\partial M}
\;  \theta ^{01} \wedge R^{23} =
 -{1\over 4 \pi^2}  \int_{\partial M}
\;  \omega ^{01} \wedge R^{23} \label{bound1} \ .
\end{equation}
At  $r_0$ this expression coincides with the contribution from the
outer
boundary of the volume integral (with the opposite sign)
and we get
\begin{equation}
 S_{GB}= S_{GB}^{volume} + S_{EGH}^{bound} =  {1\over 4
\pi^2}\left(\int_{\partial V } -
\int_{\partial M }\right)
\omega^{01}
\wedge R^{23}=-  {1\over 4 \pi^2}\left(\int
\omega^{01}
\wedge R^{23}  \right)_{r=r_h }  \ .
\end{equation}
It is easy to recognize the Chern 3-form
\begin{equation}
\Pi =   {1\over 4 \pi^2}\omega^{01}
\wedge R^{23}
\end{equation}
and the fact that the non-vanishing contribution is coming only from
the inner
boundary,
when calculating the volume integral. Before we will use the specific
form of
different black hole metrics we may summarize the result as follows.

We have shown that for a 4-dimensional compact manifold with an outer
 boundary
the G-B action is given by the integral of the Chern 3-form $\Pi$ at
the
horizon of the black hole.

Thus for our class of solutions we get
the following  action
\begin{equation}
S_{GB} = \int \Pi |_{r_h}  \ .
\label{GBgeneral}\end{equation}
If one takes into account  eq. (\ref{cond}), which is equivalent to
the
condition that the black hole has a finite and non-vanishing
temperature, one
gets
\begin{equation}
S_{GB} = \int \Pi |_{r_h} = 2   \left(1 - (e^{U} R')^2 \right)|_{r_h}
 \ .
\label{GBnonextr}\end{equation}

\section{G-B theorem for the  non-extreme black holes}

To verify the relation between the Euler
 number of the non-extreme
black holes and the G-B action  presented above we would like to
consider
different known static
axion-dilaton black holes.

i)  Non-extreme $U(1)$ dilaton black holes. The
metric is given by eq. (\ref{met}), where
\begin{equation}
e^{2U}
= (1- {r_+ \over r}) (1- {r_- \over r})
^{{1-a^2 \over 1+a^2 }}, \qquad R = r \;(1- {r_- \over r})
^{{a^2 \over 1+a^2} } \ ,
\end{equation}
where the mass and the charge of the black hole are
\begin{equation}
M= {r_+\over 2}+\left ({1-a^2 \over 1+a^2 }\right)  {r_-\over 2},
\qquad Q^2 =
{r_+ r_-
\over 1+a^2 }  \ .
\end{equation}
Non-extreme black holes have $r_+ > r_-$ and the event horizon is at
$r_+=
r_h$. For all of those black holes at the horizon $e^{U}|_{r_h}=0$
and
$R'(r)|_{r_h}$ is non-singular. Therefore the  G-B
action as given in eq. (\ref{GBnonextr}) becomes equal to 2 since
 \begin{equation}
(e^{U} R')^2
|_{r=r_h}=0  \ .
\label{correc}\end{equation}

ii) Non-extreme $U(1)\times U(1)$ axion-dilaton black holes. The
metric is
\begin{equation}
e^{2U}=  {(r- r_-) (r- r_+) \over R^2}
, \qquad R^2 = r^2 - \Upsilon^2  \ .
\end{equation}
 The mass and the electric and the magnetic charges of the black hole
are
related to $ r_+, r_-$ and to the axion-dilaton charge is $\Upsilon$.

As long as the black hole is non-extreme, we can verify that the
combination
given in eq. (\ref{correc}) vanishes, and again  the G-B
action  (\ref{GBnonextr})  equals 2.

Thus, as expected,  for non-extreme black holes the G-B theorem with
the action
given in eq.
(\ref{GBnonextr}) leads to the correct geometrical answer.

\section{G-B action for extreme black holes}

Extreme black holes have either vanishing temperature (surface
gravity at the
horizon) or  finite temperature.  We have agreed not
to consider solutions
with formally infinite temperature.
Consider first extreme black holes with  vanishing temperature
(surface
gravity).
\begin{equation}
(e^{2U})'
|_{r=r_h}= 0  \ .
\end{equation}
Such black holes usually have a regular horizon
but the range of the $\tau$ integration is infinite.  Again we cannot
apply
the G-B theorem since the space is not compact in
$\tau$-direction. One way to fix this problem would be to consider
the volume
integral to be extended only to $r=r_h + \epsilon$ simultaneously
with
restricting the range of $\tau$-integration by
$0\leq \tau  \leq  \beta$. Even if  ignored the issue of compactness
and  tried to calculate the G-B volume integral we would find that
we are in a situation where
the product of $\beta \times (e^{2U})'
|_{r=r_h} = \infty \times 0 $ is not well defined and some additional
assumption is
required. One simple procedure  which we begin by
tentatively trying out and then rejecting is to assume that, as
before, the
product is
\begin{equation}
\beta \; (e^{2U})' |_{r=r_h + \epsilon }= 4\pi\ .
\label{cond2}\end{equation}
This is in agreement with the prescription above which makes the
range of the
$\tau$-integration finite.
 If such restriction has been imposed, which
may be taken as our definition of the limit to extreme,
we still have to examine the
behavior of the term $\left((e^{U} R')^2\right)_{extr}
|_{r=r_h}$ if one first takes the limit to extreme and then
afterwards takes
the
limit of this expression as we approach the horizon.

{\it Examples}

i)  Extreme Reissner-Nordstr\"om black holes.
\begin{equation}
e^{2U}
= (1- {r_+ \over r})^2
, \qquad R = r ,
\end{equation}
and the mass and the charge of the black hole are
\begin{equation}
M=Q = r_+  \, \ .
\end{equation}
The function $R(r)$ is just $r$ and therefore
\begin{equation}
\left((e^{U} R')^2\right)_{extr}
|_{r=r_+}= 0 \ ,
\label{nocorr}\end{equation}
 and G-B action (\ref{GBgeneral}) equals $2$ under
the condition (\ref{cond2}).

ii) Extreme $U(1)\times U(1)$ black holes. Here again eq.
(\ref{nocorr}) is
valid and we see that for the extreme $U(1)\times U(1)$ axion-dilaton
black holes the  G-B action (\ref{GBgeneral})
 with eq. (\ref{cond2}) imposed equals~2.

iii)  Extreme $U(1)$ dilaton black holes with $a\leq 1$.
\begin{equation}
e^{2U}
= (1- {r_+ \over r})
^{{2 \over 1+a^2 }}\ , \qquad R = r \;(1- {r_+ \over r})
^{{a^2 \over 1+a^2} }\ ,
\label{extr}\end{equation}
where the mass and the charge of the black hole are
\begin{equation}
M= {r_+\over 1+a^2 } \ ,
\qquad Q^2 =
{(r_+) ^2\over 1+a^2 }  \ .
\end{equation}

The temperature of black holes with $a < 1$ vanishes. Therefore we
have to
repeat
the discussion of the problem in the beginning of this section with
the same
proposal to keep condition given in eq. (\ref{cond2}).
We assume again
that the integration over $\tau$ is from $0$ to $\beta$, and $\beta$
is defined
in eq.
(\ref{cond2}). The basic difference between these solutions and the
one
discussed
above is the fact that the function $R'(r)$ at the horizon is
singular. The
function $e^{U}$ still vanishes at the horizon, by the definition of
the
horizon. However, the product
$e^{U} R'$ does not vanish at the horizon anymore. The G-B action
(\ref{GBgeneral})
becomes
\begin{equation}
S_{GB} =  2 - \left(e^{U} R')^2\right)_{extr}
|_{r=r_h} = 2 -\left({a^2\over
1+a^2}\right)^2   \ .
\label{volextreme}\end{equation}
This situation is obviously non-satisfactory.\footnote{The only
exception is
the
Reissner-Nordstr\"om extreme black hole with $a=0$.}  The G-B volume
integral
with the boundary term at outer infinity, given in eq.
(\ref{GBgeneral}),
is neither
 an integer nor a topological invariant. Indeed,  we can see a strong
dependence on the metric through the dilaton coupling $a$.

One might  try to blame our prescription for dealing  with
$\tau$-integration
for
these extreme solutions. However, if we  consider the stringy case
with
$a=1$ we will
find that the problem is not due to this prescription. The reason for
that is
the following.
The surface gravity of $a=1$ extreme $U(1)$ dilaton black hole is
finite.
Therefore one does not encounter a  problem since there exists a
natural choice
of $\tau$-integration in this case, which is the same as in all
non-extreme
cases with finite temperature.
The resulting G-B action (\ref{GBgeneral}) is still not satisfactory,
\begin{equation}
S_{GB} =  2 - \left(e^{U} R')^2\right)_{extr}
|_{r=r_h} = \left[2 -\left({a^2\over
1+a^2}\right)^2\right]_{a=1} = {3 \over 2}   \ .
\label{a1}\end{equation}

To clarify the situation  we consider the domain of  integration to
be such
that
$r$ does
not reach  the horizon. We also add the boundary terms at  the
horizon.

The boundary near  the horizon $r= r_h+\epsilon$ will be defined by
the metric
in
eq. (\ref{bounmet}) with $r_0 = r_h+\epsilon$. In the end of
calculations we
will send $\epsilon$ to zero.
It is clear that we will have a complete cancellation of the volume
term
by the boundary term if we use in the boundary term the $\Pi$-form of
Chern
as suggested in \cite{EGH},
\begin{equation}
S_{GB}= S_{GB}^{volume} + S_{GB}^{boundary}=\int_{\partial V}  \Pi  -
\int_{\partial M}  \Pi  = 0\ ,
\label{zero}\end{equation}
since now
\begin{equation}
\partial V = \partial M  \ .
\end{equation}

Thus we have found that by adding the inner boundary we get complete
agreement with Gauss-Bonnet
theorem and that for all extreme black holes with singular horizons
$\chi=0$.
We  conclude that the extreme black holes are very different from the
near extreme ones despite the fact that
the extreme black holes have a small difference in the metric
as compared to the non-extreme ones when the mass  is only slightly
greater than the extreme value.

Since we were forced to make  this crucial step for all $0<a\leq 1$,
it seems
natural to
look back
 and to revise our treatment of $a=0$ extreme
Reissner-Nordstr\"om black
holes as well as $U(1)\times U(1)$ black holes. Indeed, if we stress
that
by  definition  any extreme black hole  is the one which allows
to identify the imaginary time coordinate $\tau = it$ with {\it any}
period
$\beta$, we may well  relax the  assumption about the area of
integration
in $\tau$-direction, given in eq. (\ref{cond2}). Under such condition
we could
 get
{\it any} result from the volume integral supplemented by the outer
boundary contribution. The only way to get a {\it unique} result is
to
add the inner boundary. As we have explained above, this will lead
unambiguously
to $\chi=0$ for the extreme Reissner-Nordstr\"om black
holes as well as extreme $U(1)\times U(1)$ black holes.

The calculations given above are entirely consistent with the
following
geometric picture. In all  extreme cases the Killing vector field
${\partial }\over {\partial \tau}$  vanishes at no finite point in
the
manifold.
The horizon is at infinite distance. By contrast in the non-extreme
case the
horizon is at a finite distance and must be included to complete the
manifold.
However if one does so one now has an entire 2-sphere of fixed points
at the horizon. The different values for the Euler
 number now
 follow from the general fixed point theorems described in \cite{GH}.
The difference is precisely that between a ``cigar''  and a
``pipette''
and is illustrated in Figs. 1 and 2.

\section{Entropy of extreme black holes}

If we accept the point of view, presented above, about the existence
of the
inner boundary for all extreme black holes, we have to revise the
calculation
\cite{KOP}
of the on-shell action and of the entropy for such black holes,
as was first  been emphasized by S. W. Hawking in lectures at the
Newton
Institute
describing a preliminary version of the results reported in
\cite{HHR}. A
remarkable property of $U(1)$ dilaton black holes with $0<a\leq 1$ is
the
following. The boundary conditions are not relevant for these
calculations and
 the change
in the boundary conditions does not affect the result. The
semiclassical
on-shell action as well as the entropy vanish for all these solutions
no
matter whether we introduce the inner boundary or not. To explain
this we
will repeat the calculation of the Lagrangian, for example for
extreme
dilaton electric black holes.

Our starting point for the calculation of the on-shell action will be
the following
Lagrangian
 \begin{equation} \label{action1}
 I = \frac{1}{16\pi}\int d^4x\, \sqrt{-g}\, \left({\cal L}_{grav} +
{\cal L}_{dil}  +{\cal L}_{ gauge} \right) \ .
\end{equation}
The gravitational
part of the action has a Landau-Lifshitz form
\begin{equation}\label{LL}
\sqrt{-g}\,{\cal L}_{grav} =  -\,\sqrt{-g}\  R +
\partial_\mu  \sqrt{-g}\, \omega^\mu \ ,
 \end{equation}
where the vector $\omega^\mu$ in the total derivative term in the
gravitational Lagrangian ($K$-term) is
\begin{equation}\label{eqforomega}
\omega^\mu = g^{\lambda \rho} \Gamma^{\mu}_{\lambda \rho} -
g^{\lambda \mu} \Gamma^{\nu}_{\lambda \nu}\ .
\end{equation}
The  $K$-term  removes the second derivatives
of the metric from the Lagrangian.
The gravitational part of the
action is given by eq. (\ref{LL}), and the vector $\omega^\mu$ in the
total derivative
term in the  Lagrangian is given by eq.(\ref{eqforomega}).
Eq.(\ref{eqforomega}) can be also given in the form
\begin{equation}\label{landau}
\omega^\mu = -\, \frac{1}{\sqrt{-g}}\partial_\lambda  \left(
\sqrt{-g}g^{\lambda\mu}\right)  - g^{\lambda\mu}\left(
\partial_\lambda
\ln {\sqrt{-g}}
\right) \ .
\end{equation}
For these calculations there will be no need to rewrite the volume
integral for the  total derivative part in the Lagrangian (second
term in eq.
(\ref{LL})) as a  surface integral ($K$-term). Also we will not
transform the
gauge part of  the action to a surface integral.
It will be sufficient to keep all terms in a volume integral in what
follows.

The dilaton part of the Lagrangian is
\begin{equation}
\sqrt{-g}\,{\cal L}_{dil} =  2\sqrt{-g}\
 \partial^\mu \phi \cdot \partial_\mu \phi  \ .
\end{equation}
The gauge part of the Lagrangian for the purely electric solution is
\begin{equation}
\sqrt{-g}\,{\cal L}_{ gauge} =
\sqrt{-g}\,{\cal L}_{ electr} = -\,\sqrt{-g}\
{\mbox{e}}^{-2\phi}\, F^{\mu\nu}F_{\mu\nu}\ .
\end{equation}
The maximally supersymmetric purely electric
extreme black holes are  described by the following metric \cite{US}
\begin{equation}\label{metr}
ds^{2} =
{\mbox{e}}^{2U}dt^{2}-{\mbox{e}}^{-2U}d\vec{x}^{2}\ .
\end{equation}
Before using the field equations, let us calculate the total
derivative term
in the gravitational part of the Lagrangian for the ansatz
(\ref{metr}).
We find using eq. (\ref{landau}) that
\begin{equation}
\partial_\mu  \left(\sqrt{-g}\, \omega^\mu\right) = -\, 2 \,
\partial_i
\partial_i
U\ .
\end{equation}
The total gravitational part of the Lagrangian becomes
\begin{equation}
\sqrt{-g}\,{\cal L}_{grav} = -\, \sqrt{-g}\, R \,
 -\, 2 \, \partial_i \partial_i U \ .
 \end{equation}
At this stage we may start taking the equations of motion into
account. The dilaton for maximally supersymmetric extreme black holes
is related to the metric as follows:
\begin{equation}
\phi = U \ .
\end{equation}
The first equation of motion which will be used to calculate the
on-shell
Lagrangian is the one which relates the scalar curvature to the
dilaton
contribution,
\begin{equation}
R - 2 \partial^\mu \phi \cdot \partial_\mu \phi = 0 \ .
\end{equation}
It  follows that, on-shell,
\begin{equation}
\sqrt{-g}\, \left({\cal L}_{grav} +
{\cal L}_{dil} \right) =
- 2 \, \partial_i \partial_i U\ .
\end{equation}
To treat the gauge action we have to use another equation of
motion,
\begin{equation}
\nabla^{2}\phi -
{\textstyle\frac{1}{2}} e^{-2\phi}F^{2} = 0 \ .
\end{equation}
For the electric solution with $U= \phi$ it leads to
\begin{equation}
\sqrt{-g}\,{\cal L}_{ electr} = 2 \, \partial_i \partial_i \phi\ ,
 \end{equation}
and the total on-shell Lagrangian becomes
\begin{equation}
\sqrt{-g}\,{\cal L} = -2 \, \partial_i \partial_i U + 2 \, \partial_i
\partial_i \phi =0 \ .
 \end{equation}
The same procedure allows to establish
that the on-shell Lagrangian for all $0<a\leq1$ black holes vanishes,
as well
as for
pure magnetic ones with $a=1$. This is the only known to us
configuration,
where the calculation of the entropy of extreme solutions coincides
with the limit from non-extreme ones. This can be easily seen in Fig.
3 where
the $a=1$ pure electric or pure magnetic solutions are in the
corners.
The Lagrangian
and therefore the action vanishes for $PQ=0$
dilaton black holes.  In other
words, we get the
same result by evaluating the action directly as we get by taking the
$PQ=0$
limit of the expression for black holes with regular horizon.

Note that  in the process of calculation of the on-shell Lagrangian
for maximally
supersymmetric extreme dilatonic black holes, we never faced the
problem of
going to Euclidean signature, choosing a proper gauge for the vector
potentials,
and thinking about boundary surfaces (the horizon versus infinity).
All of those problems, however, arise for the non-maximally
supersymmetric
extreme  $U(1)\times
U(1)$ dilaton black holes (Reissner-Nordstr\"om black holes can be
viewed as
a particular case of $U(1)\times
U(1)$ dilaton black holes \cite{US}). The corresponding
configurations on the
Fig. 3 correspond to the sides of the diamond.

In \cite{KOP}, the generalization of the Gibbons-Hawking method of
the calculation of the Euclidean action for dilaton black holes was
presented.
One starts with non-extreme black holes characterized by some finite
temperature
and surface gravity, and performs the calculation of the
on-shell action in Euclidean signature
by compactifying the Euclidean time coordinate. It turns out that one
can express the total action as a surface
integral and
evaluate the contribution from the extrinsic curvature and the gauge
terms
from
the outer boundary. As a final step, the extremal limit was
considered, and
the result
was:
\begin{equation}\label{A}
 S_{\chi=2} =  {\textstyle\frac{1}{4}} A = \pi (M^2 - \Sigma^2) =
{\textstyle\frac{1}{2}} \pi |z_1^2 - z_2^2| =2\pi |PQ|\ ,
\end{equation}
where $z_1$, $z_2$ are the central charges of extreme black holes
defined in \cite{US} and $P$ and $Q$ are the electric and magnetic
charges of
the 2 $U(1)$ fields. One can see this on the border of the shaded
region on
Fig. 3.
This corresponds to the topology of the near extreme Euclidean black
hole, presented in Fig. 1. If however we would like to associate with
the
extreme black holes the topology, presented in Fig. 2, this will
affect
our calculation of the Euclidean action for all extreme
configurations which
are not maximally supersymmetric, for which $|z_1^2 - z_2^2| \neq 0
$.

The idea is that if we regard the spacetime as having
an inner boundary the action will have to be augmented by an extra
boundary term and this will affect the calculation of the action and
hence the
entropy  \cite {HHR}. If we add an inner boundary for these
configurations, we
must
subtract the value of $S$ which comes from the outer boundary. This
can be
easily understood by looking at the calculation of the entropy in the
Sec. 2
of \cite{KOP}. The result is that now entropy vanishes,
\begin{equation}\label{A1}
 S_{\chi=0} =
{\textstyle\frac{1}{2}} \pi |z_1^2 - z_2^2| - {\textstyle\frac{1}{2}}
\pi |z_1^2 - z_2^2| =0 \ .
\end{equation}
 This is essentially the generalization of the result found
originally by
Hawking for $\Sigma=0$ and described
in
\cite{HHR}  and  confirmed by  Teitelboim \cite{T}. We find this
conclusion particularly
satisfying because one does not expect a soliton state to have
non-vanishing
entropy. It also fits in with our interpretation of the functional
integral
about
these backgrounds as giving a Witten Index.

\section {The Witten index}

In this section we shall discuss the relevance
of the previous results to calculations of the Witten index for
supergravity
theories.
To begin with recall that for any supersymmetric theory one may
define the
Witten index $W(\beta)$ by
\begin{equation}
W(\beta)= Tr_{\cal H}  e^{-\beta H} (-1)^F \ ,
\end{equation}
where $H$ is the Hamiltonian, $F$ the fermion number operator and the
trace
is taken over the suitable Hilbert space of theory. Note that if the
factor
$(-1)^F$ is replaced by $1$ we obtain the Gibbs partition function
$Z(\beta)$
at the temperature $T=\beta^{-1}$.

If one represents $W(\beta)$ by a formal path integral then {\it
both} the
bosonic {\it and} fermionic fields should be taken to be periodic
with period
$\beta$, in contrast to the Gibbs partition function for which  the
bosonic
fields are periodic but the fermionic fields are antiperiodic.
Formally at
least,
if the Hilbert space ${\cal H}$ consists of states which are paired
by a
supersymmetry operator $Q$ then $W(\beta)$ should be independent of
$\beta$ and its numerical value should be strictly zero. If one adds
the
vacuum state which is invariant under all supersymmetry operations
$Q$ then
the value of $W(\beta)$ will be unity.

The case we shall mainly be interested in is when ${\cal H}$ consists
of a
short supermultiplet of soliton states together with their quantum
fluctuations. The lowest, purely bosonic, soliton state is taken to
be an
extreme black hole. For our purposes an classical extreme black hole
solution
may be taken to be one which allows one to identify the imaginary
time
coordinate $\tau = i t$ with {\it any} period $\beta$. In the
semiclassical
expansion, which corresponds to large $\beta$, one may attempt to
evaluate
$W(\beta)$ by a functional integral over matter fields, including the
metric
$g_{\mu\nu}$, which are periodic in imaginary time with period
$\beta$. The
dominant contribution might be anticipated to arise from a classical
extreme
black hole solution whose mass is $M$. The main result from the old
and more
recent discussions of extreme black holes is that, even for the case
of extreme
black holes in Einstein-Maxwell theory, the classical euclidean
action $I$
works out to be given by
\begin{equation}
I = M\beta \ ,
\label{action}\end{equation}
where $M$ is the total mass. To obtain this result one must,
as we have seen, pay careful
attention to boundary condition and boundary terms. In particular one
must
treat the extreme black holes as being essentially different from
non-extreme
black holes. Thus one should not evaluate the action $I$ by taking a
limit. If
one does so one misses an essential inner boundary term not present
in the
non-extreme case. We refer the reader to  \cite{HHR}, \cite{T} for
more details. It follows from (\ref{action}) that the classical
entropy $S$ of
extreme black holes vanishes,
\begin{equation}
S=0 \ .
\end{equation}
This result has already been obtained for extreme $U(1)$dilaton black
holes
\cite{US}. It is also true that for extreme black holes, the topology
of the
Riemannian section is $S^1\times R\times S^2$ and hence that the
Euler
characteristic $\chi$ vanishes.
\begin{equation}
\chi = 0 \ .
\end{equation}
Note that, as we have seen,  if one wishes to evaluate $\chi$ using
the
Gauss-Bonnet theorem one
must pay special attention to boundary terms. In particular there is
an inner
boundary contribution in the extreme case which is absent in the
non-extreme
case.

Extreme black holes are known to be invariant under global
supersymmetry transformations, generated by Killing spinor fields
$\epsilon(x)$ which tend to constant at infinity
\begin{equation}
\lim_{ |\vec x|\rightarrow \infty} \epsilon (x) \rightarrow \epsilon
\ .
\end{equation}
In a theory with $N$ supersymmetries, $N'$ of which are unbroken the
soliton
states will fall into short supermultiplets of dimension
$2^{2(N-N')}$. For
example, in $N=2$ supergravity there are $2N=4$ possible
supersymmetry
generators. $ 2N'$ of them (corresponding to the rest frame, say)
leave the
solution
invariant and $2(N-N')= 2$ generate a  supermultiplet of 4 states
with spins
($0^+, +1/2, - 1/2, 0^-)$. Each state in the multiplet has, at the
tree level,
the
same mass. Because the entropy $S=0$ no further degeneracy is
introduced by
tree level thermodynamic effects such as are encountered for
non-extreme
black holes.

We turn now to the calculations of fluctuations around the classical
soliton
backgrounds. To calculate such fluctuations, for example in
perturbation
theory using functional determinants, one must adopt suitable
boundary
conditions. As we indicated above the appropriate choice is for the
fermions to
be periodic in imaginary time with period $\beta$. In fact, it seems
rather
plausible, though we have not calculated this directly, that if we
adopt
anti-periodic boundary conditions for the fermions, the one-loop
terms would
diverge on the horizon because of the well-known Tolman redshifting
in a
background gravitational field with $g_{00} \neq {\rm constant}$.

In fact rather than treat the fluctuations in components it is more
efficient
to
treat the fluctuations using superfield techniques as has recently
been
demonstrated by one of us \cite{K}, \cite{US} and by C. Hull
\cite{CH}. In that
work the arguments were presented that at higher than one-loop order
no
appropriate supersymmetric counterterms exists consistent with the
background supersymmetry. The basic idea is that the existence of
Killing
vectors allows to choose a coordinate system where the solution is
independent
on some bosonic coordinate of the space. The presence of a Killing
spinor in
turn implies the possibility to choose a coordinate system in
superspace such
that the solution is independent of some direction in the
anticommuting
coordinates of the superspace. Actually the argument is valid not
only for the
counterterms but for any corrections to the effective action. This
improvement
is based on the fact that the manifestly supersymmetric Feynman graph
rules\footnote{For $N=4$ theory for which manifestly supersymmetric
rules are
not
known, one may use the Feynman rules which have only $N=2$
supersymmetry
manifest. It is remarkable that for the black holes with some
unbroken $N=4$
supersymmetry, the corresponding manifestly realized $N=2$
supersymmetry
also has some unbroken part.}
 always lead to the expressions local in fermionic variables.
Therefore,
the total expressions,  describing the  Feynman supergraphs become
local in
anticommuting variables after some intermediate integrations. This
property
is sufficient to prove the vanishing of the superspace integrals over
the
whole superspace, when the integrand is independent on one or more of
the
fermionic coordinates of the superspace. This result follows directly
from
Berezin rules of integration over anticommuting variables. One has to
take into
account that the arguments, given above are formal. They correspond
to the
statement, which follows from the formal Ward Identities and are
valid in any
order of perturbation theory. It is well known from experience with
non-abelian gauge theories, that the analysis of such formal Ward
Identities is
necessary but not sufficient condition for absence of quantum
corrections. The
first place where we encounter the problem with the arguments, given
above is
related to one-loop corrections. The so-called conformal anomaly is
expressed
in terms of the integral not over the total superspace, but as the
integral
over
the chiral superspace only\footnote{ In addition, other
supersymmetric
expressions may exists, which are presented by integrals not over the
total
superspace but only part of it. Such terms require special
considerations,
since they may or may not include the integration over the Killing
directions.
Hopefully, there is a finite number of such terms as different from
the
infinite
number of generic corrections to the effective action, described by
the full
superspace integrals. We are grateful to K. Stelle for this
observation.}.

We may  study  the subtleties in supersymmetric
non-renormalization theorems related to the one-loop conformal
anomalies. For
$N=4$ supersymmetric theories such anomalies are absent in some
versions of
the theory. For
$N=2$ they exist.
The corresponding  integral over the chiral superspace does not
automatically
vanishes for the extreme Reissner-Nordstr\"om configuration. However,
the
part of the anomaly, which is local in $x$-variables is given by the
Euler
characteristic  $\chi$ of the background solution. In particular, the
only one-loop counterterm in all pure supergravities is proportional
to
the Euler characteristic $\chi$ and the coefficient is related to the
anomaly.
For a general solution of the Einstein-Maxwell equations the
divergent
counterterm does exist, since both the anomaly coefficient is
non-vanishing and
the Euler characteristic $\chi$ is non-vanishing. The vanishing of
the Euler
characteristic for extreme black hole solutions is thus a necessary
condition
for the consistency of the picture we are proposing and we find it
very
satisfying that this condition is indeed met.

The result is that the contribution to $W(\beta)$ from the spin zero
soliton
state is given, to all orders in perturbation theory, by
\begin{equation}
W(\beta) = \exp (-M \beta)\ .
\end{equation}
Note in particular the argument outlined above indicates that the
maximal
central charge condition, valid for $N=2$ supersymmetry,
\begin{equation}
M^2 = {P^2\over \kappa^2}\ ,
\end{equation}
where $\kappa^2 = 4\pi G$, $P$ is the magnetic charge and $G$ is
Newton's
constant is maintained in all orders by virtue of the fact that
neither the
mass
nor the magnetic charge receive quantum corrections. Summing over the
4 soliton states we than obtain the answer
\begin{equation}
W(\beta) = 0\ .
\end{equation}
The argument sketched above applies to the Witten index in the case
that
${\cal H}$ is the one-soliton Hilbert Space. In the next section we
will
attempt to extend it to the case of more than one soliton.

\section  {Multi-Black Holes}

In addition to classical solutions representing single isolated black
holes
there exist multi-black hole solutions representing an arbitrary
number, $k$
say, of black holes admitting Killing spinors \cite {GibKastor}.  In
what follows we shall, for completeness describe the solutions and
their moduli
spaces for arbitrary values of  the dilaton coupling constant $a$
despite the fact that only the values $ a^2 = 0, {1 \over 3}, 1 $ and
$3$
have yet been identified as arising from a supergravity theory. The
spacetime metric is
\begin{equation}
ds^2= -V^{{-2 }\over {1+a^2}} dt^2 + V^{{2 }\over {1+a^2}} d{\bf x}
^2 \ ,
\end{equation}
where $V=V( {\bf x})$ is given by
\begin{equation}
V = 1 + \Sigma ^{i=k}_{i=1} {{\mu_i} \over {| {\bf x}- {\bf x}_i |}} \ .
\end{equation}
The points $ {\bf x}_i$ correspond to extreme horizons.

Just as in the case of a single extreme black hole in the case
of the multi black holes the Euler characteristic of the
corresponding
Euclidean
solutions vanishes. The easiest way to see this is to
use the fact that the Killing vector field ${\partial} \over
{\partial \tau}$
where $\tau = it$ has no fixed points. Then the result follows
by Hopf's theorem
about everywhere non-vanishing vector fields. On a manifold with
boundary
(which may have more than one connected component) Hopf's theorem
states that the sum of the indices
of a vector field which is either transverse to the boundary or
every-where
parallel to the boundary equals to Euler number of the manifold.
In our case we consider the four-dimensional sub-manifold $N$ of
the Euclidean multi-black hole spacetime ( with $\tau$ identified
with some
arbitrary period) for which $0 < \epsilon_1 \le V ^{-1} \le
\epsilon_2 <1$ for
some  $\epsilon_1$ and $\epsilon_2$. If $\epsilon_2$ is close to $1$
then
the set of points $V ^{-1} = \epsilon_2$ may be regarded as the
component of
the
boundary near infinity. If $\epsilon_1$ is very small then the set of
points
for which $V^{-1} = \epsilon_1$ gives us an inner boundary with $k$
connected components. Since the length  of the vector field
${{\partial } \over
{\partial \tau}}$ is $V^{{-1} \over (1 + a ^2) }$ it is everywhere
non-vanishing in the sub-manifold $N$. Thus the indices vanish.
 The vector field also
clearly lies in the three-dimensional submanifolds $ V = {\rm
constant}$.
Thus it is parallel to the boundary.

Of course we could also apply the Gauss-Bonnet theorem with boundary
as we did
earlier for the case of one black hole
but the calculation would be more involved because of the more
complicated
form of the metric and the additional internal boundaries, one  for
each
horizon.

Thus just as in the case of a single extreme black hole the
counterterms
coming from the Gauss-Bonnet action must vanish on these backgrounds.
The entropy is also clearly zero. This
is completely consistent with our idea that the functional
integral corresponds to the Witten index. Now since the Witten index
is
purely topological it is
unchanged by  changes of the temperature parameter $\beta$ it should
be possible to evaluate it for very large $\beta$, i.e. at very
low energies. However the low energy  quantum behaviour of solitons
is believed to be governed by quantum mechanics on their  \lq \lq
moduli
space\rq\rq
or space of zero-modes. The Witten index should then be related to a
topological
index of the relevant quantum mechanics. Since we have seen that the
Witten
index vanishes, partly  because of the vanishing of the Euler number,
it would seem that the relevant index on the moduli
space must also vanish. In what follows we shall
investigate to what extent this is true.

To begin with  consider the geometry and topology of the moduli
spaces.
If the black holes are identical we have $\mu_1=\mu_2= \dots
=\mu_k=\mu$.
For fixed $\mu$ the solutions then depend on $3k$ parameters or
 \lq\lq moduli \rq \rq . The space of parameters is  just the \lq \lq
moduli
space\rq \rq ${\cal M} _k$. In the present case the moduli space is
given by the $k$
position vectors ${\bf x}_i$ subject to the condition that no two
coincide.
Moreover since the holes are identical, permutations of the ${\bf
x}_i$'s
give the same configuration.
Thus the moduli space corresponds to the configuration space
$C_k( {R}^3)$ of $k$ unordered distinct points in ${R}^3$, i.e.
\begin{equation}
{\cal M}_k \equiv C_k ( {R} ^3) \equiv \Big ( \Bigl ({R} \Bigr ) ^3 -
\Delta
\Bigl )/ S_k\ .
\end{equation}
where $\Delta$ is the \lq \lq diagonal set \rq \rq for which at least
two of
the ${\bf x}_i$'s
coincide and $S_k$ is the permutation group on $k$ letters.

In what follows we shall be more interested in the \lq \lq
relative moduli space \rq \rq
${\cal M}_k ^ {\rm rel} = C_k ({R}^3) / {R}^3 $. This is obtained by
identifying configurations which differ by
an overall translation. Thus
\begin{equation}
{\cal M}_k  \equiv {\cal M}_k ^ {\rm rel} \times {R} ^3\ .
\end{equation}
In the simplest non-trivial case, $k=2$, it is easy to see that the
relative
moduli space is given by the non-zero position vector
\begin{equation}
{\bf r} = {\bf x}_2 - {\bf x} _1
\end{equation}
with ${\bf r}$ and $-{\bf r}$ identified. Thus topologically
\begin{equation}
{\cal M} _2 ^ {\rm rel} = {R} \times R  P ^2\ ,
\end{equation}
where  $RP ^2$ is the
2-sphere with opposite points identified.
This is a non-orientable manifold with Euler characteristic equal to
one.
In fact this is  general, i.e.
\begin{equation}
\chi ( {\cal M} _k ^{\rm rel}) =1\ .
\end{equation}
To see that this is true  we use the Poincar\'e polynomial
$P(t,\tilde {{\cal M}_k} )$ of $ \tilde {{\cal M}_k}$,
the universal covering space of
${\cal M}_k$. This is the space obtained if we do {\sl not}
identify the points, corresponding to ${\tilde C}( R  ^3)_k$.
We have, by definition, for any manifold
\begin{equation}
P(t,{\cal M})= \Sigma ^{p= {\rm dim} {\cal M}} _ {p=0} \ \ t^p B_b (
{\cal M})\ ,
\end{equation}
where the Betti numbers
$B_p( {\cal M})= {\rm dim}\; H^p ( {\cal M}, {R})$ of the $p$'th
cohomology group of  ${\cal M}$. Clearly
\begin{equation}
\chi ({\cal M}) = P(-1, {\cal M})
\end{equation}
For $C_k({R} ^n)$, $n \ge 3$ we have \cite {Atiyah}
\begin{equation}
P(t) = ( 1 + t^2) (1 + 2t^2) \dots (1+ (k-1) t^2)\ ,
\end{equation}
whence
\begin{equation}
\chi ( {\tilde C}(R ^n)_k= k!\ .
\end{equation}

Now, by considering a simplicial decomposition
of a manifold ${\cal M}$ and lifting it to any covering space
$\tilde {\cal M}$ one sees that
\begin{equation}
\chi ( {\cal M}) = \chi (\tilde {\cal M}) / | \Gamma|\ ,
\end{equation}
where $|\Gamma|$ is the order of the group $\Gamma$
of \lq \lq deck transformations\rq \rq
such that ${\cal M} \equiv {\tilde {\cal M}} / \Gamma $.
In the present case $\Gamma = S_k$ the permutation group
on $k$ letters
and therefore
because the order of the permutation group $S_k$ is $k!$ it follows
that
\begin{equation}
\chi ({C({R} ^n)}_k)=1\ .
\end{equation}

The moduli carries a natural metric obtained by
restricting the kinetic energy
functional to the finite dimensional submanifold of the
configuration space of static solutions.
Unlike the topology the metric depends on the dilaton coupling
constant $a$.
Consider, for simplicity, the simplest case $k=2$. The metric is
given by
\cite {Shiraishimod}
\begin{equation}
ds^2 = \gamma (r) d {\bf r} ^2
\end{equation}
where
\begin{equation}
\gamma (r)= 1- (3-a^2) {M \over r} +4 ( 1 + M {{1 + a^2} \over {2r}}
) ^
{{3-a^2} \over {1 + a^2 }} -4 \ ,
\end{equation}
and $r = |{\bf r}|$.
Thus $\tilde {{\cal M} ^ {\rm rel} _2}$ is
 asymptotically flat (as $r \rightarrow \infty $). If $ a^2 \le {
1\over 3}$
this is joined by a throat to an asymptotically conical region
(as $r \rightarrow 0$). If $ a^2 ={ 1\over 3}$  there is no conical
region, just an infinitely long throat or \lq \lq drain \rq \rq. If
$ a^2 \ge { 1\over 3}$ but $ a^2 \le 3$,
including the  string case $ a^2 =1$ there is a conical
singularity at
finite distance at $r=0$. If $ a^2 =3$  the metric is flat. The
qualitative form of the relative  moduli space in the case $a^2< {1
\over 3}$
is illustrated in figure 4 as a surface of revolution obtained by
suppressing
one of the angular coordinates.
Note that the metric has been calculated assuming that the
only fields contributing to the velocity dependent forces  are
gravity,
the vector particle and the dilaton. Interestingly the critical value
of
the dilaton coupling constant, $a^2 = { 1\over 3}$  corresponds
to the reduction of 5-dimensional Einstein-Maxwell theory to
4-dimensions. It has recently been shown that the solutions in
5-dimensions,
which represent extreme black strings are complete and
every where non-singular \cite {Gibbons Horowitz Townsend}. These
5-dimensional Einstein-Maxwell solutions may also be regarded as
solutions
of  5-dimensional supergravity theory.

Quantum mechanics on these moduli spaces,  but not with reference
to supersymmetry or a Witten index has been considered by a number of
 authors \cite {Shiraishi scat}.  Because they are not K\"ahler
as one expects if they have $N=2$ supersymmetry  let alone
 HyperK\"ahler  which one expects if they have $N=4$ supersymmetry
it is not obvious how to extend these calculations to the
supersymmetric case. By contrast the moduli spaces of BPS monopoles
{\sl
are} HyperK\"ahler \cite {Atiyah Hitchin} and this plays an important
role in Sen's work on S-duality \cite {Sen}. The fact that the moduli
space of extreme Reissner-Nordstrom black holes and of
Kaluza-Klein monopoles is not HyperK\"ahler has been a long-standing
puzzle.
It
may signal a breakdown of supersymmetry.
It may equally well indicate the need to consider some alternative
form
of supersymmetric quantum mechanics on the moduli space. A suggestion
of this kind has been attributed to Witten in \cite
{Gauntlett-Harvey}.
In the case of $a^2=3$ Paul Townsend has pointed out
that one may regard the multi-solitons as being
solutions of $N=8$ supergravity with half the maximum number of
supersymmetries.
It is then reasonable to expect a moduli space with $N=8$
supersymmetries
to be not just HyperK\"ahler but flat.  This may also explain why
the neutral fivebranes have a flat moduli space \cite {Felce Samols}.

In the light of the proposal
of this paper that it is the Witten index which is important it seems
natural
to consider a {\sl topological} theory such as the de Rham complex.
Since the moduli space is non-compact however there
are apparently, in contrast to the familiar case of a closed
manifold,
 few general results relating the existence of
harmonic forms to the topology of the manifold. For that reason
we will try to construct them explicitly.
To that end we consider the simplest case $k=2$ and to look for
square-integrable
harmonic forms.
We may  trivially factor out the dependence on the centre of mass
coordinate so
we consider ${\cal M} _2^{\rm rel}$. It seems reasonable to
restrict attention to spherically symmetric harmonic forms.

The one-form
\begin{equation}
\omega _1 = { 1 \over {r^2}} {{dr }\over {\sqrt \gamma}}
\end{equation}
and its Hodge dual
\begin{equation}
\omega _2 ={} \ast \omega _1 = \sin \theta d \theta d \phi
\end{equation}
are both closed, co-closed and hence harmonic.
They will be square integrable with respect to the volume form of the
metric
$\gamma d{\bf r}^2$ provided
\begin{equation}
\int ^ \infty _ 0 {{\gamma ^{-{1 \over 2} }\over {r^2} }} dr <
\infty\ .
\end{equation}
This condition will hold as long as
\begin{equation}
a^2 < { 1 \over 3}\ .
\end{equation}
It is a theorem
that a square integrable
harmonic form on a geodesically complete   Riemannian manifold
must be both closed and co-closed \cite {Rham}.
Thus we need not have checked the closure and co-closure if
this condition holds.

Under inversion on the 2-sphere $\{r, \theta , \phi \} \rightarrow
\{ r, \pi - \theta, \phi + \pi \}$, which corresponds physically
to interchanging the two black holes,
we have
\begin{equation}
\omega _1 \rightarrow \omega _1
\end{equation}
but
\begin{equation}
\omega _2 \rightarrow - \omega _ 2\ .
\end{equation}

Thus only $\omega _1$ remains well defined on the identified space
${\cal
M}_2$.
Now
\begin{equation}
\omega _1 = d \omega_ 0
\end{equation}
where the harmonic function or zero-form $\omega _0$ is given by
\begin{equation}
\omega _0 = \int {{dr} \over  \sqrt \gamma}\ .
\end{equation}
The harmonic function $\omega _0$
is not square integrable. It is however the limit of a
square integrable one-form.  Thus from the point of view
of $L^2$ cohomology it is trivial.
It is clear that $\omega _2$ is not the exterior
derivative of a smooth one-form.  This indicates that the square
integrable
or $L^2$ cohomology of $\{ \tilde {\cal M}_2 , ds^2 \}$
 the covering space equipped with its metric $ds^2$ is, in the
case that $a^2 < { 1 \over 3}$, given entirely by forms on the
2-sphere.
If we pass to the identified space ${\cal M}_2$ we find that the two
form
$\omega _2$ does not descend because it is odd under the antipodal
map.

 It would seem to follow from this that if we regard the black holes
as
 indistinguishable bosons then the $L^2$ cohomology of the moduli
space is
trivial
in the case $a^2 < { 1 \over 3}$. Therefore if  for example in the
case of
Einstein-Maxwell
theory (i.e. $a^2=0$) our identification of the Witten index with the
index of the de Rham complex is correct then the Witten index should
vanish
both for single extreme holes as well as for multi-extreme-black
holes.
The case $a^2 =3$ is also cohomologically trivial. The case $a^2 =1$
is
is unclear because of the conical singularity on the moduli space.

\section{Supersymmetric Sigma Model with  Black Hole
Target Space}
As the first step in direction to investigate the supersymmetric
quantum
mechanics on moduli space of extreme dilaton black holes we will
study here
the
supersymmetric sigma models in the black hole target space.
Some four-dimensional
 extreme  $a=1$ dilaton black holes  with electromagnetic fields
\cite{GM} have
been
recently reinterpreted as solutions with unbroken supersymmetry of
the
 effective
action of heterotic  string theory in critical dimension.  Generic
space-time
supersymmetric dilaton black hole is related to a supersymmetric
sigma model
with
torsion when the fundamental dilaton is not constant.  Indeed, we are
interested in
space-time supersymmetric solutions of heterotic string theory in
ten-dimensional
target space, which contains in particular
the dilatino $\lambda$ transformation rule.
The supersymmetry  rules  are  (our notation for this section are
given in
\cite{KO2}).
\begin{eqnarray}
\delta \psi_\mu &=& \bigl (\partial_\mu - {1\over 4}
  \Omega_{\mu +}{}^{ab}\gamma_{ab}\bigr)
\epsilon \ ,
   \label{eq:susy1}\\
\delta\lambda &=& \bigl (\gamma^\mu\partial_\mu\phi + {1\over 4}
                               H_{\mu\nu\rho}
\gamma^{\mu\nu\rho}\bigr)\epsilon \ ,
                              \label{eq:susy2}\\
\delta\chi &=& - {1\over 4} F_{\mu\nu}\gamma^{\mu\nu}\epsilon\ .
\label{susy}\end{eqnarray}
One can see from eq. (\ref{eq:susy2}) that bosonic configuration with
vanishing dilatino can have also vanishing supersymmetry variations
of
dilatino
$\delta\lambda(\epsilon) =0,
\epsilon
\neq 0$. However,  in presence of  dilaton,
depending on three-dimensional space coordinate
$\vec x$ one has to have a non-vanishing \footnote{One
can
avoid torsion for the dilaton depending on some null coordinate
$\gamma^u
\partial_u \phi\; \epsilon = 0$, with Killing spinors satisfying
$\gamma^u
 \epsilon = 0$, like in the case of gravitational waves.} torsion $H$.
 Thus, as long as we are considering dilaton supersymmetric black
holes with non-trivial dilaton field, we have geometries with torsion
from the point of view of supersymmetric sigma model. It is
interesting that if we
look on solution with
constant dilaton, we may afford to have unbroken space-time
supersymmetry
without
torsion. For example one may choose the metric to be self-dual
stringy
ALE solutions \cite{Bian}. This solution  indeed forms a target space
of the
supersymmetric sigma model with  HyperK\"ahler geometry. Stringy
multi-monopole solutions \cite{Kh} which are T-dual to a special
class of
ALE solutions, formed by multi-center metrics of Gibbons-Hawking
type,
do have a non-trivial dilaton as well as non-vanishing torsion and
therefore the
geometry is not HyperK\"ahler. Still multi-monopoles provide a target
space for
a  sigma model with world-sheet  $(4,4)$ supersymmetry \cite{CHS}
which is
non-HyperK\"ahler target space.

Magnetic black holes
were uplifted in \cite{NEL}, \cite{KO2}. The details of
supersymmetric
uplifting are described in \cite{BKO2}, \cite{BKO3}. When
supersymmetric
solution of four-dimensional  $N=4$ supergravity is reinterpreted as
a
solution of a heterotic string effective action with unbroken
supersymmetry
it always relates the four-dimensional vector field of the charged
$a=1$
black hole to a non-diagonal component of the metric, which is equal
to a
2-form field
$B_{\mu 4}$.  The supersymmetric sigma model of such theory always
has
torsion and the geometry can not be HyperK\"ahler!  In some cases
T-duality
transformation  can be performed on the target space  as well as on
world-sheet variables
 which is consistent with supersymmetry and removes torsion. However,
generic  space-time supersymmetric dilaton $a=1$ black hole before
any
duality transformation is related to a supersymmetric sigma model
with
torsion when the fundamental dilaton is not constant. More details
about
supersymmetry versus T-duality can be found in \cite{BKO3}.

The $(4,0)$ supersymmetric sigma  model for magnetic uplifted dilaton
black
hole \cite{GM}  was discussed  in \cite{NEL} and presented explicitly for the
throat limit. We will present here the $(4,1)$ supersymmetric sigma
model for
magnetic non-Abelian uplifted dilaton black hole without considering the
throat limit.

It was explained in
\cite{KO2}
 that the magnetic black hole solution  has to  be supplemented by
the
non-Abelian vector field  to avoid Lorentz and supersymmetry anomaly.
The
uplifted magnetic black hole geometry corresponds to a decomposition
of the
manifold
$M^{1,9}
\rightarrow M^{1,5} \times M^4$ and the tangent space
$SO(1,9)\rightarrow
SO(1,5) \times SO(4)$.  The curved manifold is  four-dimensional
$M^{4}$,
it has Euclidean signature and it has a self-dual curvature for a
torsionful
connection. The six-dimensional manifold  $ M^{1,5}$ is flat.
Four-dimensional manifolds are known to have interesting properties
when
considered as target space for supersymmetric sigma models.  The
supersymmetric sigma models $(p, q)$ where $p $ is the number of left
and
$q$ is the number of  right supersymmetries,  are described in detail
in
\cite{WEST} for  $ p, q  \leq 2$. An extensive  list of references
can also be
found there. In addition to that for our purpose it will be
sufficient to use
the description of supersymmetric sigma models with $p$ or $q$ equal
$4$
as given by Hull and Witten \cite{HW} and Howe and Papadopolous
\cite{HOWE}. Our method of investigation of the world-sheet
supersymmetry
of the sigma model in the black hole target space will be very close
to the
method which was applied by Callan, Harvey and Strominger  \cite{CHS}
for
the analysis of the world-sheet supersymmetry $(4,4)$ of the sigma
model
in the target space of the  fivebrane.  In what follows we are going
to show
that the non-Abelian magnetic black hole target space provides
$(4,1)$
supersymmetry of the sigma model. Note that we are suggesting a world-sheet
description of the magnetic dilaton black holes which differs from the one
used in \cite{GPS}. The main difference is  in the role which the
the vector field plays. For us the four-dimensional abelian $U(1)$ vector
field becomes a
non-diagonal component
of the metric as well as of the two-form field in a compactified dimension
$x^4$. In addition we have a non-Abelian $SU(2)$ vector which will appear in
the  interaction with the left-handed world-sheet fermions.

 The generic $(4,1)$ supersymmetric sigma model \cite{HOWE} with
manifestly
realized $(1,1)$ supersymmetry in the four-dimensional Euclidean
space is
given by
\begin{equation}
I_{(4,1)} = \int d^2 z d^2 \theta  (G_{\mu\nu} + B_{\mu\nu}) D_+
X^\mu D_-
X^\nu\  ,
\label{superaction}\end{equation}
where the unconstrained $(1,1)$ superfield is given by
\begin{equation}
X^\mu (x^\mu, \theta^+, \theta^-) = x^\mu (z) + \theta^+
\lambda_+{}^a (z)
e_a{} ^\mu (x)
 - \theta^-
 \lambda_-{}^a(z)   e_a{}^\mu (x)   - \theta^+\theta^- F^\mu (z)\ .
\label{superfield}\end{equation}
Action (\ref{superaction}) has additional  3 supersymmetries for some
special
backgrounds
$G_{\mu\nu}(X) ,  \;  B_{\mu\nu}(X)$
which admit covariantly constant complex structures $J_r {}^\mu{}_\nu
= e_a{}
^\mu \;J_r {}^a{}_b\;
e^b {} _\nu $
with special properties. Let us show that the non-Abelian uplifted
magnetic
black hole \cite{KO2}
\begin{eqnarray}
ds^{2}_{(10)} & = & dt^2 - e^{4\phi}d\vec{x}^{2}-(dx^{\underline 4} +
V_{\underline i} \, d x^{\underline i})^{2}-
dx^{\underline I}dx^{\underline I}\, ,
\nonumber \\
\nonumber \\
B_{(10)} &= &  {1\over 2} B_{\mu\nu} \; dx^\mu \wedge dx^\nu  = - V_
{\underline
i}\,  d x^{\underline i} \wedge dx^{\underline 4}
\, ,
\nonumber \\
\nonumber \\
\partial_ {\underline i }  \partial_{\underline i }  e^{2\phi} & = &
0\, ,
\qquad
e^{2\phi}  =  1+ \sum_s \frac{2M_s}  {|\vec x - \vec x_s|}\, .
\qquad
2 \partial_{[ \underline i} V_{\underline k ]} =  \pm  \epsilon
_{ikl}
\;\partial_{\underline l }
e^{2\phi}\, \nonumber \\
\nonumber \\
 A_{k}{}^ {i m} &=&  \Omega_{k - }{}^{im} =
2
\delta_{k[i}   \partial _ {\underline m]} e^{-2\phi}
\qquad
A_4^{ij} = \Omega_{4 - }{}^{ij} =  \pm \epsilon _{ijk}
\partial
_{\underline k}
e^{-2\phi}\ .
\label{upl}\end{eqnarray}
has all the corresponding properties. This solution has flat six
dimensions
and non-flat
four-dimensional Euclidean space $(\vec x, \; x^4)$. The vierbein
basis  is
$e_{\underline i}
{}^j = e^{
2\phi}\delta _{\underline i}{} ^j , \;
e_{\underline i}{} ^4= V_{\underline i}, \; e_{\underline 4}{}^i = 0,
\;
e_{\underline 4}{}^4=1$.
Thus, one can use the
general form of the $(1,1)$ supersymmetric action (\ref{superaction})
 with
the non-trivial four-dimensional part of the metric and 2-form field
as
given in (\ref{upl}) and $\mu, \nu  ; a,b = 1,2,3,4$.
\begin{equation}
G_{ij} (X) = - e^{4\phi (X)}\delta_{ij} - V_i (X) V_j  (X) , \quad
G_{i4}
(X) = B_{i4}(X) = -V_{i}(X)\ ,
\quad G_{44}=-1
\end{equation}
where the functions $\phi$ and $V_i$ of the $(1,1)$ superfield $X$
are the
same as the corresponding functions  of $x$ in (\ref{upl}).  Let us
remind
the reader that the non-Abelian vector field was obtained via spin
embedding
into the gauge group and therefore it is a quantity derived from the
given
values of the metric (vierbein) and a 2-form field.
Also the second torsionful connection as well as both curvatures are
derived quantities: if one has the metric and the 2-form field, one
can
derive other characteristics of the configuration. In the sigma model
the
same property is seen as follows.
If one would like to perform the integration over the world-sheet
fermionic
coordinates
one would recover the supersymmetric action, in which however $(1,1)$
supersymmetry is not realized manifestly anymore. The action  in
addition
to the first term, depending on ordinary $x$, will contain terms with
covariant derivatives acting on right-moving fermions $ \lambda_+{}^a
$ and
left moving fermions $ \lambda_-{}^a $ as
well as the term quartic in fermions. In the covariant derivatives of
the
right(left)-moving fermions one has to use the torsionful
spin-connection
$ \Omega_{\mu + }{}^{ab}\; (\Omega_{\mu - }{}^{ab})$. The quartic
term
contains the torsionful curvature which has  the  exchange properties
for
the  positive and negative torsionful spin connections under
condition $dH=0$.

The origin of additional 3 world-sheet supersymmetries is related
directly
to the fact
that our black hole has space-time unbroken supersymmetries
\cite{KO2}.
Eqs. (\ref{susy}) have a zero mode with
\begin{equation}
(1 \pm \gamma_5)  \epsilon
_{\pm} \equiv (1 \pm \gamma_1\gamma_2\gamma_3\gamma_4) \epsilon
_{\pm} = 0\ .
\label{constr}\end{equation}
For one choice of the sign in solution one gets a constant spinors
with
positive chirality $\epsilon_+$ , for the other choice one gets a
constant
negative chirality spinor $\epsilon_-$.

The difference between black hole target space and that of the
fivebrane
\cite{CHS} at this stage is the following. For the fivebrane one of
the
torsionful spin connection is self-dual and the other one is
anti-self-dual.
This means that for one choice of sign in the solution one has
simultaneously
\begin{equation}
 \bigl (\partial_\mu - {1\over 4}
  \Omega_{\mu +}{}^{ab}\gamma_{ab}\bigr)
\epsilon _{+} = 0\ , \quad   \bigl (\partial_\mu - {1\over 4}
\Omega_{\mu -}{}^{ab}\gamma_{ab}\bigr)
\epsilon _{-} = 0\ .
\end{equation}

Therefore
one promotes   $(1,1)$ supersymmetric sigma model to the $(4,4)$
 supersymmetric one. The procedure is described  in \cite{CHS}. One
constructs 3 self-dual complex structures out of space time Killing
spinors
of one type of chirality and the 3 anti-self-dual complex forms out
of the
spinors of the opposite chirality.  For the black hole target space
we have
only $ \bigl (\partial_\mu - {1\over 4}
  \Omega_{\mu +}{}^{ab}\gamma_{ab}\bigr)
\epsilon _{+} = 0$ from the gravitino part of unbroken space-time
supersymmetry and therefore the self-dual properties of $\Omega_+$
spin
connection.
\begin{equation}
\Omega_{+i}{}^{jk} \;  \epsilon _{jkl} =  2 \Omega_{+i}{}^{4l}= -2
\epsilon_{ikl} \partial_{\underline k}e^{-2\phi} \ .
\end{equation}
The other property of the fivebrane $  \bigl (\partial_\mu - {1\over
4}
\Omega_{\mu -}{}^{ab}\gamma_{ab}\bigr)
\epsilon _{-} = 0$ is not valid for the black hole (\ref{upl}). The
$\Omega_-$ spin connection is presented in eq. (\ref{upl}) and it is
clear
that it it is not (anti)self-dual. Thus we can construct only one
type of 3
complex structures, which allows to promote the world-sheet
supersymmetry
to $(4,1)$ only. The reason for that is the following. One of the
properties of the 3 complex structures $J_r{} ^a{}_b$ is  that they
have to be
covariantly
constant with respect to either $ \bigl (\partial_\mu - {1\over 4}
  \Omega_{\mu +}{}^{ab}\gamma_{ab}\bigr)
$ for promoting to $(4,1)$ or with respect to $\bigl (\partial_\mu -
{1\over 4}
\Omega_{\mu -}{}^{ab}\gamma_{ab}\bigr)
$ to promote supersymmetry to $(1,4)$. In the black hole case only
the
first possibility is available. Thus we conclude that the
supersymmetric
sigma model action in the black hole target space has $(4,1)$
world-sheet
supersymmetry\footnote{Our analysis does not exclude that there are
more
supersymmetries which may still be hidden, however $(4,1)$ is
certainly
available.}.  The black hole provides an excellent example of  a
constrained extended $(4,1)$ superfield $X^\mu( x^\mu, \theta^+_0,
\theta
^+_r, \theta^-), r=1,2,3$, satisfying the following constraint
\cite{HOWE}
\begin{equation}
D_{r +}  X^\mu = J_{r \;\nu}^\mu
D_{0+}  X^\nu\ .
\end{equation}
Three complex structures $J_{r \;\nu}^\mu $ are constructed from the
Killing
spinors of unbroken space-time supersymmetry of the black hole
configuration. They satisfy all conditions required by Howe and
Papadopolous \cite{HOWE} for the $(4,1)$ supersymmetric sigma model.
In
particular, the complex structures obey the quaternion algebra, they
are
covariantly constant with respect to torsionful
$\Omega_+$-connection and the holonomy group is $Sp(1)$. However, the
manifold is obviously not HyperK\"ahler since there is a torsion $H
_{4ij}
= \pm
{1\over 3} \epsilon_{ijk} \partial _{\underline k} e^{2\phi}$.

The relation of supersymmetric sigma model in the target space given
by
uplifted
magnetic black hole manifold to the geometry of the moduli space
still has to be understood. However, we believe that the sigma model
presented above clearly indicates that there is no reason for the
$a=1$
dilaton black holes to rely on HyperK\"ahler manifold for  $N=4$
supersymmetric quantum mechanics. Rather one may expect the manifold
of the
type described above where extended supersymmetry of quantum
fluctuation
exists due to the existence of unbroken space-time supersymmetry on
manifold with torsion.

\section{Towards $N=4$ supersymmetric Quantum Mechanics  of Black Holes}

To confirm the conclusion of the previous section we may dimensionally
reduce the  supersymmetric sigma model on the world sheet to the
one-dimensional theory on the world line.
It is useful for this purpose  to perform the integration over the fermionic
coordinates
in (\ref{superaction}) and eliminate auxiliary fields. This will also
permit us to  compare our  dimensionally reduced action on black hole
manifolds with  the supersymmetric quantum mechanics of monopoles in
$N=4$ Yang-Mills  theory \cite{Blum}.  One gets the following
Lagrangian
\begin{eqnarray}
{\cal L}_2&=&  (G_{\mu\nu} + B_{\mu\nu}) \partial_z x^\mu \partial_{\bar z}
x^\nu +i \lambda_+{}^a (\nabla_{z}{}^{(+)} \lambda_+) ^a - i \lambda_-{}^a
(\nabla_{\bar z}{}^{(-)} \lambda_-) ^a \nonumber\\
\nonumber\\
&-& {1\over 2} R^{(+)}_{ab,cd} \lambda_+{}^a  \lambda_+{}^b
\lambda_-{}^c
\lambda_-{}^d \ .
\label{lagr2}\end{eqnarray}
The right(left)-handed fermions $\lambda_+{}^a$ ($\lambda_+{}^a$)
have covariant derivatives
with respect to torsionful spin connections $\Omega_{+}$ ($\Omega_{-}$).
The torsionful curvatures $R_{\pm} = d\Omega_{\pm} + \Omega_{\pm}\wedge
\Omega_{\pm}$ have the exchange properties
\begin{equation}
R^{(+)}_{ab,cd}= R^{(-)}_{cd,ab} \ .
\label{exch}\end{equation}
Those exchange properties generalize the symmetry properties of the torsionless
Riemann tensor $R_{ab,cd}= R_{cd,ab}$.
 For our non-Abelian black hole the torsionful curvatures are given either by
the Yang-Mills field strength or by the gravitational torsionful curvature,
depending on the position of the indices. One is expressed through another by
eq.
(\ref{exch}). Because of that one could rewrite the last term in (\ref{lagr2})
as
\begin{equation}
{1\over 2} R^{(+)}_{ab,cd} \lambda_+{}^a  \lambda_+{}^b
\lambda_-{}^c
\lambda_-{}^d = {1\over 4} R^{(+)}_{ab,cd} \lambda_+{}^a  \lambda_+{}^b
\lambda_-{}^c
\lambda_-{}^d + {1\over 4} R^{(-)}_{ab,cd} \lambda_-{}^a  \lambda_-{}^b
\lambda_+{}^c
\lambda_+{}^d \ .
\end{equation}
We would like to stress that  for the black holes $R^{(+)}_{ab,cd}\neq
R^{(+)}_{cd,ab}$ as well as
$R^{(-)}_{ab,cd}\neq R^{(-)}_{cd,ab}$  which is another manifestation of the
fact that we deal with manifolds with torsion.

Lagrangian (\ref{lagr2}) has $(4,1)$ supersymmetry. Four
right-handed ones relate $x$ with $ \lambda_+$ and one left-handed relates
$x$ with $ \lambda_-$, as explained above in terms of the action
depending on superfields. Lagrangian (\ref{lagr2}) is invariant up to terms
which are total derivatives in $z$ or $\bar z$. Therefore if we will perform
dimensional reduction of the Lagrangian (\ref{lagr2}) by requiring $x,
\lambda_+, \lambda_-$ to be independent on $\sigma $ and dependent on $\tau$
we will keep all abovementioned supersymmetries of the Lagrangian preserved.
The reduced lagrangian is
\begin{eqnarray}
{\cal L}^{QM}_{bh}&=& {1\over 2} G_{\mu\nu}  \partial_\tau x^\mu \partial_\tau
x^\nu +{i\over \sqrt 2}\lambda_+{}^a (\nabla_{\tau}{}^{(+)} \lambda_+) ^a -
{i\over \sqrt 2}
\lambda_-{}^a  (\nabla_{\tau}{}^{(-)} \lambda_-) ^a \nonumber\\
\nonumber\\
&-& {1\over 4} R^{(+)}_{ab,cd} \lambda_+{}^a  \lambda_+{}^b
\lambda_-{}^c
\lambda_-{}^d - {1\over 4} R^{(-)}_{ab,cd} \lambda_-{}^a  \lambda_-{}^b
\lambda_+{}^c
\lambda_+{}^d \ .
\label{QM}\end{eqnarray}
The torsion is gone from the bosonic part of the action, since the term
$B_{\mu\nu} \partial_\tau x^\mu \partial_\sigma x^\nu$ is not available
anymore. However, the fermionic part of one-dimensional action (\ref{QM})
keeps track of torsion, it remains in both covariant derivatives as well as in
the
fact that there are two different curvature tensors. Covariant derivatives on
fermions are
\begin{eqnarray}
(\nabla_{\tau}{}^{(+)} \lambda_+) ^a &\equiv& \partial_\tau \lambda_+{}^a +
\partial_\tau x^\mu\;\Omega_{+\mu }{}^{ab} \lambda_{+b}\nonumber\\
\nonumber\\
(\nabla_{\tau}{}^{(-)} \lambda_-) ^a &\equiv& \partial_\tau \lambda_-{}^a +
\partial_\tau x^\mu\; \Omega_{-\mu }{}^{ab} \lambda_{-b} \ .
\end{eqnarray}
In notation of \cite{KO2} we have $\Omega_{\pm} = \omega \mp {3\over 2} H$,
where $\omega$ is metric compatible spin connection and the torsion for the
black hole manifold is given by the
3-form
\begin{equation}
H = \pm dx^4\wedge dx^i \wedge dx^j \; \epsilon _{ijk}\;
 \partial _{\underline k} \; e^{2\phi} \ .
\end{equation}
which is closed since the dilaton $e^{2\phi}$ satisfies harmonic equation of
motion $\partial_{\underline k}\partial_{\underline k}  e^{2\phi}= 0$
\begin{equation}
dH = \mp dx^4\wedge dx^i \wedge dx^j \wedge dx^l \; \epsilon _{ijk}\;
\partial_{\underline l} \partial _{\underline k} \; e^{2\phi}=0  \ .
\end{equation}
The last condition $dH=0$ is necessary for the proof of the exchange properties
of
curvatures (\ref{exch})  and for the proof of supersymmetry of the action
(\ref{QM}).

Thus we have got an one-dimensional supersymmetric action which has at least
$N=4$ supersymmetry or more and codifies nicely all information about the
uplifted non-Abelian magnetic $a=1$ extreme dilaton black hole (\ref{upl})
whose manifold is not HyperK\"ahler. This action certainly
represents some possibility to work out the supersymmetric quantum mechanics
related to the black hole. The  quantization of this one-dimensional action
most
likely will not lead to any problems and we hope to perform it and report about
the results elsewhere.

It is rather instructive to compare our action (\ref{QM}) with the one,
associated
with the
the supersymmetric quantum mechanics of monopoles in $N=4$ Yang-Mills
theory  in the form given in \cite{Blum}.
\begin{eqnarray}
{\cal L}^{QM} _{mon}= {1\over 2} G_{\mu\nu}  \partial_\tau x^\mu
\partial_\tau x^\nu  -  {i\over 2}
\bar \lambda^a  \gamma_0 (\nabla_{\tau} \lambda) ^a - {1\over 2} R_{ab,cd} \bar
\lambda ^a  \bar \lambda^b \lambda^c \lambda^d \ .
\end{eqnarray}
This one-dimensional supersymmetric action is a corollary of  the analysis
of the boson and fermion zero modes in the monopole moduli space and it is
supposed to describe the supersymmetric quantum mechanics of the monopoles
in their moduli space. The metric here is HyperK\"ahler, the covariant
derivative
of fermions is   metric compatible and there is one torsionless Riemann tensor
in the four-fermion coupling.

In our case the one-dimensional supersymmetric action (\ref{QM}) is available
and looks rather unique from the point of view of the possibilities to have any
$N=4$ supersymmetric quantum mechanics, related to the black hole geometry
(\ref{upl}). The relation of this action to the moduli space approach as well
as the quantization of this theory remain to be investigated.

\section{Fission of extreme black holes}
In this section we explore approach to extreme black holes as a
special quantum
system. Having revised our view on the entropy of all extreme black
holes we
are in  a better position to discuss and develop the idea, suggested
by   Linde
and described in the Appendix of  \cite{US} about the splitting of
the extreme
black holes.  It was suggested there that the probability of this
process is suppressed by the factor $e^{\Delta S}$, where $\Delta S$
is the change of the total entropy of the system of black holes.

The simplest example of splitting concerns $U(1)$ $a=1$ purely
magnetic
(or purely electric) extreme black holes. They have vanishing entropy
(and vanishing area of horizon). Therefore they presumably can split
(or diffuse quantum mechanically, without any energy release)
into smaller purely magnetic or electric black holes, since
$\Delta S =0$ in this case.

 A more complicated example considered in \cite{KOP} was
 the splitting of the extreme $U(1)\times
U(1)$
electric-magnetic black hole into a purely magnetic and a purely
electric one. At the time that this example was studied it was
qualified as
follows:
such bifurcation  is forbidden classically but  could in principle
occur in a quantum-mechanical process. The reason for this is that
classically,
if cosmic censorship holds, and appropriate energy conditions
then a theorem of Hawking states
that black holes cannot bifurcate and their area cannot decrease.
Setting aside the purely {\sl classical} concept
of cosmic censorship, if one identifies one quarter of the area of
the event
 horizon with entropy then  the
fact
that the
entropy of extreme $U(1)\times
U(1)$
electric-magnetic black hole was previously considered to be $S=2\pi
|PQ|$ as
explained
above and that the entropy of the products of the splitting was
vanishing
according to the same formula, it would also follow that splitting
was forbidden {\sl thermodynamically}. It seems to us that while one
might
be prepared to give up cosmic censorship in the quantum or near
quantum regime,
and while one can certainly not expect the energy conditions to hold
in
all circumstances in the quantum regime one {\sl would}  expect
the results of thermodynamics to remain true.
 Now that we have come to conclusion that the entropy of all
extreme black
holes vanishes, the splitting  of black holes discussed above  starts
with
zero entropy state and ends with zero entropy state. Such processes
can be
viewed as  decays of nuclei or  elementary particles. We will present
some
details of this splitting here since this is the simplest one and we
will later
compare it with the black hole splitting in the theory with dilaton
coupling
$\sqrt 3$ where the process  of splitting one black hole into two
will be
accompanied by the release of energy.

Thus we consider first the decay of  one  extreme $a=1$ $U(1)\times
U(1)$
electric-magnetic black hole into pure electric $U(1)$ and pure
magnetic
$U(1)$.  It
can be described by
\begin{equation}
(P, Q) \rightarrow (P,0) + (0, Q) \ .
\end{equation}

The extreme
electric-magnetic black hole has the following relations between
parameters:
\begin{equation}
M_{initial}= \frac {|P|+ |Q|}{\sqrt{2}} \ ,  \hskip 2 cm \Sigma =
\frac {|P|-
|Q|}{\sqrt{2}}\ ,
\end{equation}
which gives
\begin{equation}
M^2 + \Sigma^2 - P^2 - Q^2 = 0 \ .
\end{equation}
For simplicity consider the situation with both charges positive
and $P \geq Q$. For this case the solution above has one quarter of
supersymmetries of $N=4$ supergravity unbroken, which corresponds to
$\epsilon _+^{34}$ and $\epsilon ^+_{34}$, see ref. \cite{US} for
details.
 The
parameters of the daughter black holes are related to those of the
parent as
\begin{eqnarray}
\label{multi}
M_{final} &=& M_1 + M_2 \ ,\hskip 2 cm M_1 = \Sigma_1 = \frac
{|P|}{\sqrt{2}} \ ,\nonumber\\ \nonumber\\
\Sigma&=& \Sigma_1 + \Sigma_2 \ , \hskip 2 cm
M_2 = -\Sigma_2 = \frac {|Q|}{\sqrt{2}}\ .
\end{eqnarray}
and
\begin{eqnarray}
M_1^2 + \Sigma_1^2 - P^2  &=& 0 \ ,\nonumber\\
\nonumber\\
M_2^2 + \Sigma_2^2  - Q^2 &=& 0 \ .
\end{eqnarray}

For positive charges this means that the magnetic solution by itself
has
 one half of unbroken supersymmetries $\epsilon _+^{34}, \; \epsilon
^+_{34}$ and
$\epsilon _-^{12}, \; \epsilon ^-_{12}$. The electric solution by
itself has  one half of unbroken supersymmetries $\epsilon _+^{34},
\;
\epsilon ^+_{34}$ and
$\epsilon _+^{12}, \; \epsilon ^+_{12}$ \cite{US}. However,
 everywhere in
space the total configuration has the same unbroken supersymmetry as
the
parent solution, i.e. one quarter of $N=4$ unbroken supersymmetries
given by
$\epsilon _+^{34}, \; \epsilon
^+_{34}$.

These black holes are in an equilibrium with each other, since the
attractive
force between them vanishes due to supersymmetry \cite{US}.
Indeed, let us consider Newtonian,
Coulomb and
dilatonic forces.  The force between two distant objects of masses
and charges
$(M_1, Q_1, P_1, \Sigma_1)$ and $(M_2, Q_2, P_2,\Sigma_2)$ is
\begin{equation}
F_{12} = - \frac{M_1 M_2}{r_{12}^2} +
\frac{Q_1 Q_2}{r_{12}^2} +
\frac{P_1 P_2}{r_{12}^2} - \frac{\Sigma_1 \Sigma_2}{r_{12}^2} \ .
\end{equation}
The dilatonic force is attractive for charges of the same sign and
repulsive for charges of opposite sign.
Using the relations (\ref{multi}) for the masses and dilaton charges
 in terms of the magnetic and electric charges $P_1 = P$, $P_2 = 0$,
$Q_1 = 0$, $Q_2 = Q$, we
see  that $F_{12}$ vanishes.

There is no release of energy during this splitting since
$M_{initial}- M_{final}=0$
but it is also not suppressed
since there is no change in entropy, according to our new picture.

Now consider the new case, where indeed we deal with fission.
Consider the theory
\begin{equation}
S = {1\over 16 \pi} \int d^4x\,\sqrt{-g}\left( R +2\partial^\mu
\phi\cdot\partial_\mu \phi
-e^{-2a\phi}  F_{\mu\nu}  F^{\mu\nu} \right) \ .
\label{a}\end{equation}
with arbitrary dilaton coupling $a$. The usual extreme
Reissner-Nordstr\"om black hole with $|P|=|Q|$ and vanishing dilaton
solves
equation of motion of this theory (\ref{a}) for arbitrary $a$. For
$a=0$ this
solution has one
half of unbroken supersymmetry when embedded into $N=2$ supergravity
\cite{GibHull}. For $a=1$ an extreme Reissner-Nordstr\"om black hole
with
 $|P|=|Q|$ and vanishing dilaton solves the equations of motion of the
theory with two
vector fields and has one quarter of unbroken supersymmetries
when embedded into $N=4$ supergravity \cite{US}.
 The solution which was called a
 $[U(1)]^2$ black holes
(to stress that one vector is electric and another one is magnetic)
also
solves  the equation of motion for the axion field, which shows that
the axion is constant.  However, it is actually  a
solution of the system of fields with only one vector field, if we do
not care about
having
$ F * F =0$. Special case of $\sqrt 2 |Q|=\sqrt 2 |P|=
M$ and
$\Sigma=0$  just means that
$ F^2=0$ and $\phi =0$.

The nice property of the solution where the dilaton is vanishing
in this
non-trivial way, is that it solves equations of motion of the theory
(\ref{a})
for arbitrary values of $a$. The dilaton eq.
\begin{equation}
\nabla^2  \phi -{1\over 2} a  F^2 =0
\end{equation}
is  solved for $\phi = const$ since $ F^2=0$. The gravitational and
vector equations
are not  affected by the value of $a$ since
$e^{-2a\phi}=1$.  Thus we have established
that our
no dilaton $\sqrt 2 |Q|= \sqrt  2 |P|=  M$ Einstein-Maxwell dyon
solves the
eqs. of motion
of the bosonic Lagrangian (\ref{a}).
Now let us focus on the $a=\sqrt 3$ case. This can be embedded into
dimensionally
reduced 5d supergravity with the dilaton being $g_{55}$. The
solutions with
unbroken supersymmetry of this theory have to saturate the following
bound
\cite{GibKastor}
\begin{equation}
M\geq {1\over \sqrt
{1+a^2} } \sqrt {Q^2 + P^2} = {1\over 2} \sqrt {Q^2 + P^2} \ .
\end{equation}

Our Einstein-Maxwell dyon with $\sqrt 2\; |Q|= \sqrt  2\;  |P|=  M$
does not
saturate
the bound. Indeed for $ |Q|= |P|=$ the bound is
\begin{equation}
M\geq  {1\over \sqrt 2} |Q| \ ,
\end{equation}
whereas the extreme Einstein-Maxwell dyon state of the same theory
has the mass
$M_{dyon} =
\sqrt 2 |Q|$.
Thus there is a gap between the supersymmetric state with
$M_{susy} = {1\over 2} \sqrt {Q^2 + P^2}$ and
$M_{dyon} =
\sqrt {Q^2 + P^2}$. We can use the fact that the extreme
Einstein-Maxwell dyon
state has the energy higher than the supersymmetric ground state of
this theory.

We consider the decay of  one  extreme  Einstein-Maxwell $ |Q|= |P|$
dyon into the supersymmetric monopole and supersymmetric electropole
of
$a=\sqrt 3$ theory.

It
can be described by
\begin{equation}
(P, Q) \rightarrow (P,0) + (0, Q) \ .
\end{equation}

The extreme
electric-magnetic black hole has the following relations between
parameters:
\begin{equation}
M_{initial} = \frac {|P|+ |Q|}{\sqrt{2}}= {\sqrt{2}} |Q| \ ,  \hskip
2 cm \Sigma = \frac
{|P|- |Q|}{\sqrt{2}}=0\ ,
\end{equation}
which gives
\begin{equation}
M^2 + \Sigma^2 - P^2 - Q^2 = 0 \ .
\end{equation}
 The
parameters of the daughter black holes are related to those of the
parent as \cite{GibPer}
\begin{eqnarray}
M_{final} &=& M_1 + M_2 = |Q| \ ,\hskip 2 cm M_1 = {1\over \sqrt 3}\;
\Sigma_1 =
\frac {|P|}{2} \ , \nonumber\\ \nonumber\\
\Sigma&=& \Sigma_1 + \Sigma_2 =0\ , \hskip 2.5 cm
M_2 = -{1\over \sqrt 3} \; \Sigma_2 = \frac {|Q|}{2}\ .
\end{eqnarray}
and
\begin{eqnarray}
M_1^2 + \Sigma_1^2 - P^2  &=& 0 \ ,\nonumber\\
\nonumber\\
M_2^2 + \Sigma_2^2  - Q^2 &=& 0 \  .
\end{eqnarray}

The magnetic black hole by itself and the electric
one by itself each have unbroken supersymmetry \cite{GibKastor}.
However the
configuration with both monopole and electropole most likely does not
have
common Killing spinors. This has to be investigated. If indeed this
turns out to be
the case, it would be very close to what we have observed during the
decay of the
$a=1$ dyon. Monopole far away from electropole has an increased
number of
unbroken supersymmetries and the same for the electropole. However
the
additional Killing spinors for monopole are different from those for
the
electropole. Therefore the total number of unbroken supersymmetries
of the
configuration after decay remains the same as before decay.

The gap in energy between the initial 1-black-hole state and final
2-black-hole state is
\begin{equation}
M_{initial}- M_{final} = ({\sqrt{2}} -1) |Q| \ .
\end{equation}
The efficiency of the energy release is
\begin{equation}
{M_{initial}- M_{final}\over  M_{initial}} = (1- {1\over \sqrt{2}} )  \sim
0.3 \ .
\end{equation}

The black holes which are the product of decay are not  in an
equilibrium
with
each other, since the attractive gravitational
force between them is overcompensated by the repelling dilaton force.
Consider again eq. (\ref{bal}) for $Q_1 = 0$ and $P_2 = 0$.
\begin{equation}\label{bal}
F_{12} = - \frac{M_1 M_2}{r_{12}^2}  - \frac{\Sigma_1
\Sigma_2}{r_{12}^2} \ .
\end{equation}
Using the relations  for the masses and dilaton charges
 in terms of the magnetic and electric charges  we
see  that
\begin{equation}
F_{12} =  - \frac{M_1 M_2}{r_{12}^2} + a^2 \frac{M_1 M_2}{r_{12}^2}=
\frac{Q^2}{2 r_{12}^2} \ .
\end{equation}
The monopole and electropole repel strongly in this
theory\footnote{In theories where the dilaton is massive the
phenomenon
of repelling of black holes after the splitting was observed in
\cite{HorHor}.}
 since the monopole and electropole have dilaton charges of the
opposite sign.
 Interesting
enough, the original state in this example was such that there was no
dilaton
anywhere. After the decay the monopole fragment carries away a
dilaton charge and
the electropole fragment carries away an exactly the same value but
the opposite
sign
 dilaton charge. The total process looks like the inverse
annihilation of
the dilaton charge.
\begin{equation}
\Sigma_{monopole}= - \Sigma_{electropole} =
 {\sqrt 3 |P|\over 2} = {\sqrt 3 |Q|\over 2} \ .
\end{equation}
It is energetically advantageous to create   opposite dilaton charges
in this
theory and to separate simultaneously the electric solution from the magnetic
one. In our next
example we will
consider the  state before fission which has an arbitrary dilaton
charge. Still
we will see that the release of energy during the fission  proceeds
by
the fragmentation of the original state with  dilaton charge
imbalance:
\begin{equation}
\Sigma_{monopole}= - \Sigma_{electropole} +
 {\sqrt 3 |P|\over 2} - {\sqrt 3 |Q|\over 2} \ .
\end{equation}

The fission described above actually will proceed  starting  with
more
general extreme electro-magnetic solution of $a=\sqrt 3$ theory as
quoted
in \cite{GibWil}.
The extreme solution in isotropic coordinates is given in terms of   two
independent
parameters, electric and magnetic charges, $Q$ and $P$ or in terms of their
special combinations $\alpha,\; \beta$.
\begin{equation}
ds^2 =-e^{2U} dt^2 + e^{-2U} d \vec x^2 \ ,
\end{equation}
where
the metric and the scalar field are given by
\begin{equation}
e^{-4U}  = CD  \ , \qquad  \exp ( - { { 4 \phi} \over { \sqrt 3}}) = { D \over
C} \ .
\end{equation}
The simplest form of the vector field is given in terms of  the magnetic
potential and dual magnetic potential $\tilde A_\phi $.
\begin{equation}
A_\phi =  P \cos \theta  \ ,  \qquad  \tilde  A_\phi =  Q \cos \theta \ .
\end{equation}
The vector field part of  the action, as usual for dilaton black holes can be
written as
$ F {}^\star  \tilde F$. Equation of motion require $d\tilde F=0$ and the
existence of the dual magnetic potential.

Functions $C,\; D$ are given by
\begin{equation}
D = \left ( 1+{ \beta \over r } \right) ^2  + {{\beta^2 (\beta + \alpha)} \over
{r^2 ( \beta - \alpha )} }  \ ,
 \end {equation}
and
\begin{equation}
C = \left ( 1-{ \alpha \over r } \right) ^2  -{{\alpha^2 (\beta + \alpha)}
\over {r^2 ( \beta - \alpha )} }  \ .
\end{equation}
The mass and the scalar charge are  given by
\begin{equation}
M = { 1 \over 4} ( \beta -\alpha) \ ,
\qquad \Sigma = { {\sqrt 3} \over 4} ( \alpha + \beta) \  .
\end{equation}
The electric and magnetic charges are
\begin{equation}
Q^2 = { \beta^3  \over 4 {( \beta - \alpha)} } \ , \qquad
P^2 = {\alpha^3  \over 4  ( \alpha - \beta )} \ .
\end{equation}
Thus
\begin{equation}
M^2 + \Sigma ^2 - P^2 -Q ^2 = 0 \ .
\end{equation}

Solution is manifestly invariant under duality transformation
\begin{equation}
P \rightarrow Q \ , \quad \phi  \rightarrow  - \phi \ , \quad  \Sigma
\rightarrow
 -\Sigma \ ,  \quad  \alpha \rightarrow  - \beta  \ .
\end{equation}

The mass of the  extreme solutions is then  given in terms of
electric and
magnetic charges by

\begin{equation}
M = { 1 \over 2 } ( P^ { 2 \over 3} + Q ^ { 2 \over 3} ) ^ {3 \over
2} \ .
\end{equation}

The electric charge of this solution does not have to be equal to the
magnetic one. There is also a non-constant dilaton now. Only if we
use a
particular case of this solution with $|P|=|Q|$ will we  get the
previously
discussed case of the Einstein-Maxwell dyon with
\begin{equation}
M = { 1 \over 2 } ( P^ { 2 \over 3} + Q ^ { 2 \over 3} ) ^ {3 \over
2}= \sqrt 2 |Q| \ .
\end{equation}
This generic extreme solution is not supersymmetric unless either
$P=0$
or $Q=0$.
It will also undergo an explosive fission into monopole
$M_1 = {1\over \sqrt 3}\; \Sigma_1 = \frac {|P|}{2}$ and electropole
with
 $M_2 = -{1\over \sqrt 3} \; \Sigma_2 = \frac {|Q|}{2}$. This time
the
 repelling force will be
\begin{equation}
F_{12} =  - \frac{M_1 M_2}{r_{12}^2} + a^2 \frac{M_1 M_2}{r_{12}^2}=
\frac{|PQ|}{2 r_{12}^2} \ .
\end{equation}

The existence of this solution
gives a richer picture of fission. The initial state has arbitrary
relation between the electric and magnetic charge and a non-vanishing
dilaton. The products of decay
are monopole and electropoles with some magnetic and electric charges
which also are not equal to each other.

One may compare the general case  $P^2 = - \left ( {\alpha \over \beta}\right)
^3 Q^2  $  with the
one
discussed above. The gap between $(P,Q)$ state and $(P,0), (0,Q)$
state is
\begin{equation}
M_{initial}- M_{final} = { 1 \over 2 } ( P^ { 2 \over 3} + Q ^ { 2
\over 3} )
^ {3 \over 2}
- {1\over 2}  (|P|+|Q|) \ .
\end{equation}
The efficiency of the energy release \begin{equation}
{M_{initial}- M_{final}\over  M_{initial}} =  1 - {  (|P|+|Q|)\over
( P^ { 2 \over 3} + Q ^ { 2 \over 3} ) ^ {3 \over 2}}
\end{equation}
is a function of $ \left ( {\alpha \over \beta}\right)$.
The maximum is reached
when the solution becomes a familiar Reissner-Nordstr\"om
one at $\alpha = -  \beta $. The maximum  was found before to be
about 0.3.

One can consider the situation with the electric charge which is
proportional to some elementary electric charge $e$ times an integer
and the
magnetic charge such that the Dirac quantization condition is
imposed.
\begin{equation}
Q= e n \ , \quad P = { m \over 2 e}\ ,  \quad PQ = { nm\over 2} \ ,
\end{equation}
where $n,m$ are some integers.
The spectrum of excited states of the $(P,Q)$ black hole will look
like
\begin{equation}
M (m,n) = { 1 \over 2 } \left ( \Bigl({ m\over 2 e}\Bigr)^ { 2 \over 3} +
\Bigl(en\Bigr) ^ { 2 \over 3} \right) ^ {3 \over 2} \ .
\label{levels}\end{equation}

Any such state with arbitrary integers $(m,n)$ will eventually
decay into the black holes with charges
$(P = { m \over 2 e} , 0)$, and  $(0, Q= e n)$ with the
release of energy, described above. A process of further splitting without
the release of energy might then  continue until $m$ elementary monopoles
$(P = {1  \over 2 e},0)$ and
$n$ elementary electropoles $(0, Q= e )$ are produced. The system
(\ref{levels}) has many discrete levels and transitions between
levels may also be possible, mostly with release of energy.

One can as well consider the situation that from the beginning we have a
$[U(1)]^2$ black hole as in \cite{US}. The electric charge belongs to one
$U(1)$ group and the magnetic one to the other. In such case  the quantization
of charges may be different.

The splitting of atomic nucleus resulting in the release of large
amounts of
energy is called fission.  In what follows we are going to estimate
the amount of energy released during the black hole decay. Suppose that the
black hole has a Planckian mass $M_p \sim 1.2\times 10^{19}$ GeV $\sim 2\times
10^{-5}$ g.  In the maximum efficiency case described above
 the energy released during the black hole fission is about $6 \times 10^{15}$
erg $\sim 120$  kg  of TNT. To compare the efficiency of black hole fission
with that of Uranium, suppose that   we   have not just one
Planckian scale black hole of $2\times10^{-5}$g but many black
holes, with a total mass 1  kg. The fission of one
 kilogram of extreme black holes will result in energy release of about 6
megatons of
TNT, to be
compared with 20 kilotons of TNT during the explosion of one  kilogram  of
${}^{235}U$.
The gravitational black hole fission bomb can be 300 times more
efficient than the atomic one!

\section{Conclusion}

We have performed the calculation of the Euler number of the
axion-dilaton black holes. For all non-extreme solutions, which
(in Minkowski space) have
regular horizons, the calculation of the  Euclidean Gauss-Bonnet
volume integral with the outer boundary terms always results in the
value $\chi = 2$ as expected from the
fact that the topology of such manifolds is $R^2 \times S^2$. The
corresponding configurations are presented in Fig. 1.

Extreme dilaton $U(1)$ black holes do not have regular horizons, at
least
according to the canonical choice of spacetime metric. Our strategy
has been firstly to calculate the Euclidean Gauss-Bonnet
volume integral using  the condition $\beta \kappa= 2 \pi$. Note that
the
temperature of these black holes $T= {\kappa \over 2\pi} = {1 \over
4\pi r_+}\;
\left({ r_+ -
r_- \over r_+ } \right)^{{1-a^2
\over 1+a^2}}$ is zero for $a<1$ or finite
for $a=1$.

Despite the fact that the horizon of extreme solutions is singular,
 the volume
integral with the outer boundary term turned out
 to be finite, but the  result was $2-
\left({a^2\over 1+a^2}\right)^2$. It is neither an integer
 nor a topological invariant, since there is an
obvious dependence on the metric.  It was
natural therefore to remove the singular horizon and to add an inner
boundary
 near the horizon. We have found that if we use the correct boundary
term as
given in
\cite{EGH}, the volume term is completely cancelled by the boundary
term.
This may be
interpreted as  discontinuity in the value of the Euler
 number
for extreme ($\chi=0$, see Fig. 2)  and non-extreme ($\chi=2$, see
Fig. 1)
black holes.

Extreme Reissner-Nordstrom and $U(1)\times U(1)$ black holes  have
regular horizons. The calculation of the Euclidean Gauss-Bonnet
volume integral  with the outer boundary terms results in the value
$\chi = 2$. One can obtain this result  by
imposing the condition that the integration over the Euclidean time
is
performed in the region from zero till $\beta$ where  the combination
$\beta \kappa$ is held fixed and equal to $2\pi$.
We should stress, however,
that this calculation has an intrinsic ambiguity, related to the fact
that the
temperature of such black holes vanishes, $T=0$. Therefore the volume
integral
is not well defined. Imposing the inner
boundary for extreme black holes seems to be  the only consistent way
to be in
agreement with the general  definition that extreme black holes are
such
as to allow one to identify  the imaginary time coordinate $\tau =
it$ with
{\it any} period~$\beta$.

The study of the topological invariants of the extreme black holes
performed
in this paper,  shows the  crucial importance of the boundary
corrections to
the Euler
 number of such geometries. We believe that at present when
the creation of such black holes in some background vector fields
seems to be
possible \cite{GIBB,Fay}, it is useful to understand the topological
properties
of these new
black holes in addition to their thermodynamical properties.

 As mentioned above, the  topology of extreme Reissner-Nordstrom black
holes as well as their entropy was studied recently  \cite{HHR,T}.
 Discussions of this issue at the Durham conference
``Quantum Concepts of Space and Time'' have shown that there is a
complete agreement about the existence of the inner
boundary for all extreme black holes, which leads  to the vanishing
Euler number as well as the entropy.

A careful analysis of the boundary conditions for the extreme
black holes
in the path integral used in this paper in the context of
Gauss-Bonnet theorem has been recently developed further.  We have
come to
the conclusion that the supersymmetric non-renormalization theorem
for
extreme black holes with unbroken supersymmetries implies the
absence
of quantum corrections to a properly defined Witten Index. It is
essential for
this
interpretation that fermion fields are taken to be {\sl periodic}
in imaginary time. This is {\sl only} possible for extreme
holes. For non-extreme holes the fermion fields must be taken to
be {\sl anti-periodic} in imaginary time, consistent with the
functional
integral giving a thermodynamic partition function.

In an attempt to study the possible approaches to supersymmetric
quantum
mechanics of the fluctuations in the extreme black hole background we
have
studied a supersymmetric sigma model in the black hole target space.
The non-K\"ahlerian character of the corresponding geometry was
explained by
the existence of the torsion.

The idea   of possible splitting and joining of extreme black
holes suggested in  \cite{US} acquires an
additional  support with the new understanding that the entropy of
all (and
not only
$U(1)$ dilaton extreme black holes) vanishes. The process of
splitting and
joining of various extreme black holes is not forbidden anymore by
the second
law of black hole physics.  A discussion of this appears in \cite
{HHR}. We have studied some examples of fission of extreme
dilaton
black holes which are energetically advantageous. The existence of a
gap between the states in a
specific model indicates  that fission might be possible with an explosive
release of energy.

To conclude, there have been several developments recently in the
theory of
charged black holes. A  new picture seems to be emerging: the
relation
between extreme and non-extreme black holes
 resembles the one between the massless $m=0$ and massive
$m>0$ non-Abelian gauge fields.
Even a very light massive vector field has 3 degrees of freedom and
the
corresponding
field theory is non-renormalizable. The massless field had 2 degrees
of
 freedom and the corresponding field theory is renormalizable. The
limit
$m \rightarrow 0$ does not coincide with the massless theory $m=0$.
Extreme black holes with the mass strictly equal to the central
charge
$M=|Z|$ are very different from the non-extreme ones with $M>|Z|$ and
their limit $M \rightarrow |Z|$ is in some important respects very
different
from the extreme case.

As with massless Yang-Mills fields,  extreme black holes seem to have
features
which open some hope to make progress in understanding the underlying
quantum theory.

\section*{Acknowledgements}
We are grateful to E. Bergshoeff, Ya. Eliashberg,  A. Hanson, S.
Hawking, G.
Horowitz,  A.~Linde, T. Ort\'\i n, M. Perry, G. Segal, A. Strominger,
L. Susskind, C. Teitelboim and R. Wagoner for   valuable discussions.  We would
like to thank D. Linde for the very efficient
program which allowed the calculation of the Gauss-Bonnet Lagrangian
for
arbitrary metrics.  R.K.  would like to express her gratitude for the
hospitality at the Newton
Institute and for  financial support.  The work of R.K. was also
supported by NSF grant PHY-8612280.

\vfill
\newpage

\section*{Appendix}
We have started the  calculation of the G-B Lagrangian
\begin{equation}
L_{GB}= R_{\mu\nu\lambda \delta} R^{\mu\nu\lambda \delta}
-4 R_{\mu\nu} R^{\mu\nu} + R^2
\end{equation}
 for the
class of metrics $ds^2 =
  e^{2U}(r) d\tau^2 + e^{-2U}(r) dr^2 + R^2 (r) d^2 \Omega$
 by using   {\it Mathematica}. The Lagrangian
was obtained in the form
\begin{equation}
L_{GB}= {4\over R^2}\,\left([(e^{2U})'\, R']^2 - (e^{2U})'' +
       (e^{2U})\,(R')^2\,(e^{2U})'' +
       2\,(e^{2U})\,(e^{2U})'\,R'(r)\,R'' \right)   \ .
\label{dmitri}\end{equation}
In particular for extreme dilaton black holes with arbitrary coupling
$a$
presented in eq. (\ref{extr}) the G-B Lagrangian was obtained in the
following form:
\begin{eqnarray}
\sqrt g\; L_{GB}= &-&{{2\,r_+ \,{{\left(1 - {{r_+ }\over r} \right)
}^
        {-1 + {2\over {1 + {a^2}}}}}}\over {\left(1 + {a^2} \right)
\,{r^2}}}\nonumber\\
\nonumber\\
&+&
  {{2\,r_+ \,{{\left(1 - {{r_+ }\over r} \right) }^
        {-1 + {4\over {1 + {a^2}}}}}\,
      {{\left({{\left(1 - {{r_+ }\over r} \right) }^
             {{{{a^2}}\over {1 + {a^2}}}}} +
           {{{a^2}\,r_+ \,{{\left(1 - {{r_+ }\over r} \right) }^
                 {-1 + {{{a^2}}\over {1 + {a^2}}}}}}\over
             {\left(1 + {a^2} \right) \,r}} \right) }^2}}\over
    {\left(1 + {a^2} \right) \,{r^2}}}  \ ,
\end{eqnarray}
which can be simplified to
 the
following form:
\begin{equation}
\sqrt g\; L_{GB}=  - {{2\,r_+ \,{{\left(1 - {{r_+ }\over r} \right)
}^{-1 + {2\over {1 + {a^2}}}}}}\over {\left(1 + {a^2} \right)
\,{r^2}}}
  \left(1 -  \left( \left(1 - {{r_+ }\over r} \right) + {{a^2}\,r_+ \over
             {\left(1 + {a^2} \right) \,r}} \right)^2 \right) \ .
\end{equation}

However, it was   easy to observe looking on eq.  (\ref{dmitri}) that
$\sqrt
g\; L_{GB}$ can be rewritten
in the form
\begin{equation}
\sqrt g\; L_{GB}= -4 {\partial \over \partial r}\left(
(e^{2U})' (1 - (e^{U} R')^2)\right) \ ,
\end{equation}
since we expected to find some total derivative. In this form we have
presented
it in eq. (\ref{append}) of the paper.

Let us show how this expression can be used for Schwarzschild
 black hole where $ e^{2U}= 1 - 2m/r$ and $R(r) = r$:
\begin{eqnarray}
\chi &= &  {1\over 32 \pi^2}\left[\int
d^4 x \sqrt{-g} \;(R_{\mu\nu\lambda \delta} R^{\mu\nu\lambda \delta}
-4 R_{\mu\nu} R^{\mu\nu} + R^2)
\right] + S_{bound} \nonumber\\
 &= &{-4\pi \over 32 \pi^2} \int_0^{8\pi m} d\tau  \int_{2m}^{r_0} dr
4 {\partial \over \partial r}{2m\over r^2} (1 - (1 - 2m/r)) - {16m^3
\over
r_0^3}= 2  \ .
\label{schw}\end{eqnarray}

The simple form of the G-B Lagrangian (\ref{append}) was confirmed
when we
have used the vierbein formulation and differential forms, as
explained in Sec. 5. For completeness we present here all the
non-vanishing
curvature forms:
\begin{eqnarray}
R^{01} &=& {1\over 2} (e^{U})'' dr d\tau \ ,
 \qquad R^{02} = - {1\over 2} (e^{2U})'  e^{U} R'  d\tau d \theta \ ,
 \nonumber\\
 \nonumber\\
 R^{03} &=& - {1\over 2}  (e^{2U})'  e^{U} R'  \sin \theta dr d\phi \
,
\qquad  R^{31} = {1\over 2} (e^{U} R')'  \sin \theta dr d\phi \ ,
 \nonumber\\
 \nonumber\\
R^{32} &=& - \sin \theta \left(1 - (e^{U} R')^2 \right) d\theta d\phi
\ ,
  \quad  R^{21} = {1\over 2}  (e^{U}
R')'  dr d\theta   \ .
\quad  \end{eqnarray}

\ifx\nopictures Y\else  \input epsf  \fi

\begin{figure}

\ifx\nopictures Y\else \centerline{\hskip 5cm \epsfbox{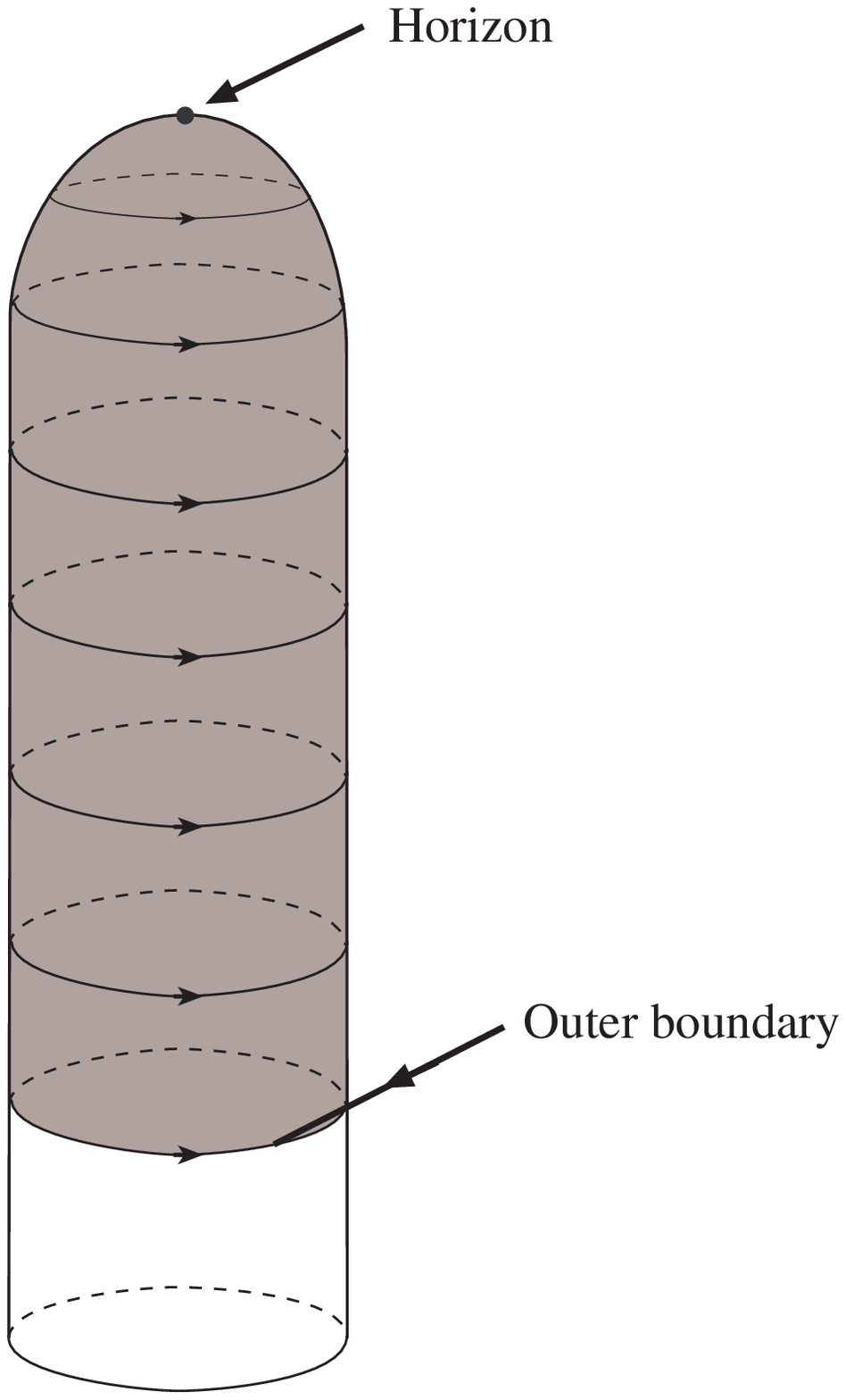}} \fi
 \vskip 1cm
\caption{The geometry of the $r-\tau$ space for non-extreme black
holes.
The circles are lines of constant $r$. The shaded region has an outer
boundary
but no inner boundary. The vector field ${\partial \over \partial
\tau}$ has a
fixed point set at the horizon. The space has the topology ${ R}^2$.
Thus
$\chi=1$.}

\label{F1}

\end{figure}

\begin{figure}

\ifx\nopictures Y\else{\centerline{\hskip 5cm \epsfbox{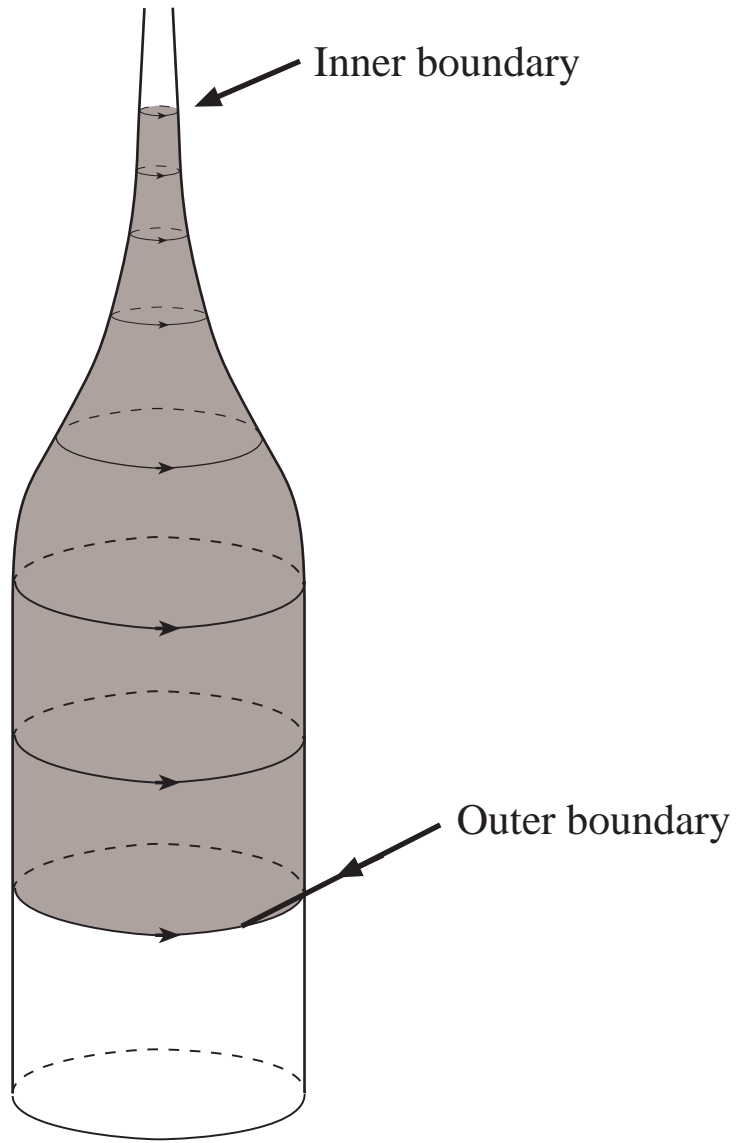}}}\fi
\vskip 1cm
\caption{The geometry of the $r-\tau$ space for the extreme black
holes.
The circles are lines of constant $r$. The shaded region has an outer
and inner boundary. The vector field ${\partial \over \partial \tau}$
has no
fixed points. The surface has an infinitely long spine, and has
topology $S^1 \times{ R} \sim R^2 - \{ 0 \}$. Thus $\chi = 0$.}
\label{F2}

\end{figure}

\begin{figure}

\ifx\nopictures Y\else{\centerline{ \epsfbox{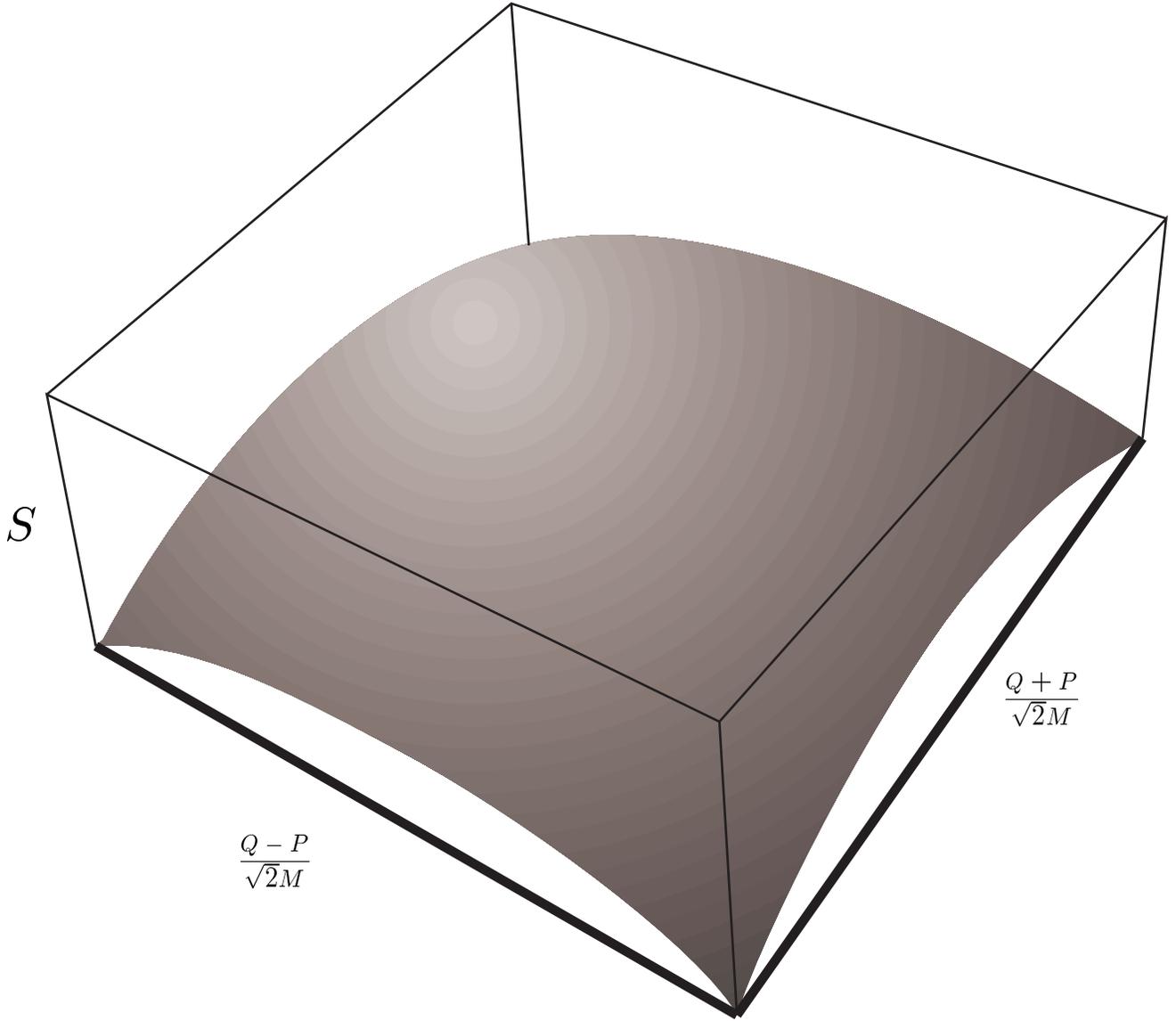}}}\fi
\vskip 1cm
\caption{The entropy $S$ of the dilaton black holes as the function
of the
electric $Q$ and magnetic $P$ charges. The shaded region presents the
entropy
of non-extreme and near extreme $[U(1)]^2$ black holes, whose
topology is shown on Fig. 1. The bold line shows the vanishing
entropy of extreme black holes, whose
topology is shown on Fig. 2.}
\label{F3}

\end{figure}

\begin{figure}

\ifx\nopictures Y\else{\centerline{ \epsfbox{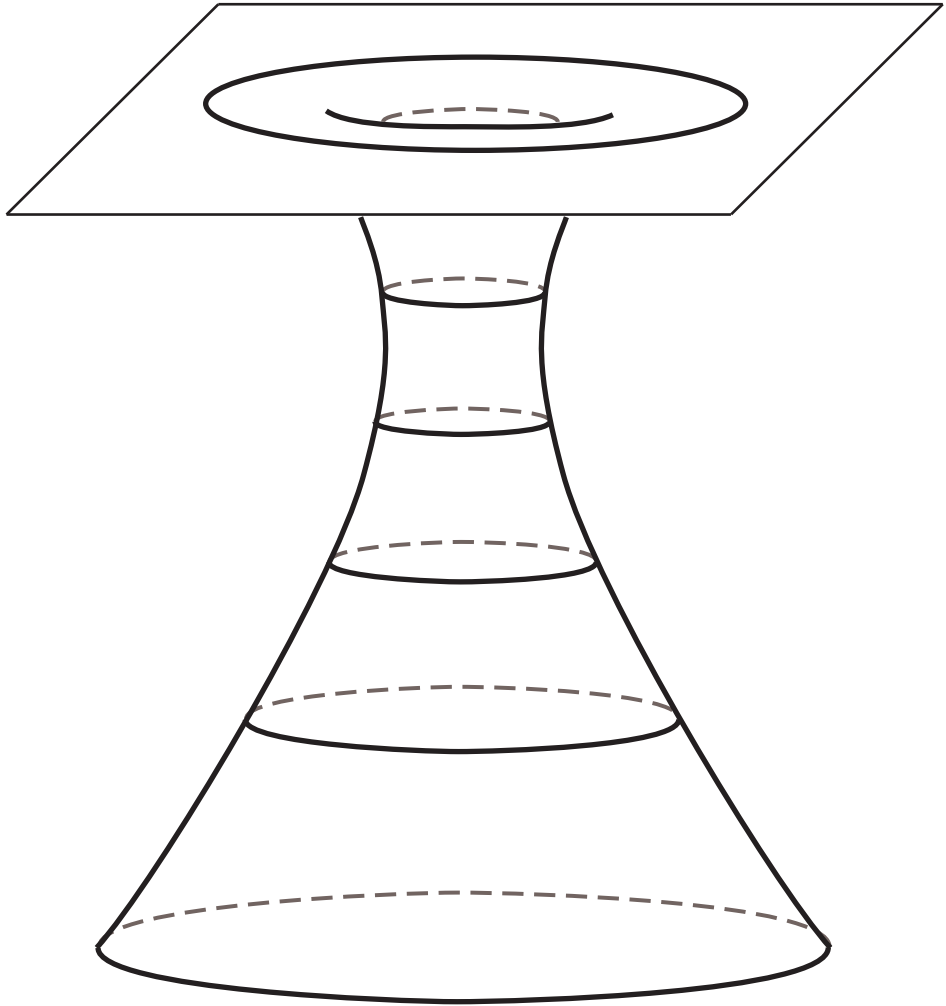}}}\fi
\vskip 1cm
\caption{The relative moduli space for two extreme holes with $0\leq
a <
{1\over 3}$.
What is illustrated is the covering space $ \tilde {{\cal M}}^{\rm
rel} _2$.
Each circle represents a
2-sphere. To obtain   ${\cal M} ^ {\rm rel} _2$ one must identify
antipodal points on these spheres. The identified space is
non-orientable and
has $\chi = 1$}
\end{figure}

\end{document}